\documentclass[manuscript]{acmart}

\usepackage{amsfonts}
\usepackage{algorithmic}
\usepackage{textcomp}
\usepackage{xcolor}
\usepackage{multirow}
\usepackage{bm}

\AtBeginDocument{%
  \providecommand\BibTeX{{%
    \normalfont B\kern-0.5em{\scshape i\kern-0.25em b}\kern-0.8em\TeX}}}
    
\newcommand{\reviewaccepted}[1]{{\textcolor{black}{#1}}}
\newcommand{\rebuttal}[1]{{\textcolor{black}{#1}}}

\begin{document}

\setlength{\intextsep}{0.cm}
\setlength{\textfloatsep}{1.2ex}
\setlength{\floatsep}{1.5ex}
\setlength{\abovecaptionskip}{0.cm}
\setlength{\belowcaptionskip}{0.cm}

\title{CNN-based Robust Sound Source Localization with SRP-PHAT for the Extreme Edge}

\author{Jun Yin}
\email{jun.yin@esat.kuleuven.be}
\author{Marian Verhelst}
\email{marian.verhelst@esat.kuleuven.be}
\affiliation{%
  \institution{ESAT-MICAS KU Leuven}
  \city{Leuven}
  \country{Belgium}}

\begin{abstract}

Robust sound source localization for environments with noise and reverberation are increasingly exploiting deep neural networks fed with various acoustic features. Yet, state-of-the-art research mainly focuses on optimizing algorithmic accuracy, resulting in huge models preventing edge-device deployment. The edge, however, urges for real-time low-footprint acoustic reasoning for applications such as hearing aids and robot interactions.
Hence, we set off from a robust CNN-based model using SRP-PHAT features, Cross3D \cite{diaz2020robust}, to pursue an efficient yet compact model architecture for the extreme edge.
For both the SRP feature representation and neural network, we propose respectively our scalable LC-SRP-Edge and Cross3D-Edge algorithms which are optimized towards lower hardware overhead.
LC-SRP-Edge halves the complexity and on-chip memory overhead for the sinc interpolation compared to the original LC-SRP \cite{dietzen2020low}.
Over multiple SRP resolution cases, Cross3D-Edge saves 10.32$\sim$73.71\% computational complexity and 59.77$\sim$94.66\% neural network weights against the Cross3D baseline.
\rebuttal{In terms of the accuracy-efficiency tradeoff, the most balanced version (\textbf{EM}) requires only 127.1 MFLOPS computation, 3.71 MByte/s bandwidth, and 0.821 MByte on-chip memory in total, while still retaining competitiveness in state-of-the-art accuracy comparisons. It achieves 8.59 ms/frame end-to-end latency on a Rasberry Pi 4B, which is 7.26x faster than the corresponding baseline. }

\end{abstract}


\settopmatter{printacmref=false}
\setcopyright{none}
\renewcommand\footnotetextcopyrightpermission[1]{}
\pagestyle{plain}

\maketitle

\section{Introduction}\label{section_introduction}
Sound source localization (SSL) \reviewaccepted{targets to derive} 
the relative position of sound sources against the origin, which is typically the recording device. 
The most recent research \reviewaccepted{focuses on} 
the calculation of the Direction of Arrival (DoA) towards the microphone array, \textit{i.e.} the source's relative azimuth and elevation angles. 
In the past few decades, SSL techniques have been exploited with various types of sound, such as ocean acoustics \cite{niu2017source}, ultrasonic signals \cite{kundu2014acoustic}, and anisotropic-material conduction \cite{kundu2012acoustic}.
Nowadays, SSL on human-audible sounds is emerging rapidly for public or domestic usage, for instance in speech recognition \cite{davila2018enhanced}, speech enhancement \cite{xenaki2018sound}, noise control \cite{chiariotti2019acoustic}, and robotic perception \cite{rascon2017localization}. 

There has been a long history of solving SSL problems with conventional signal processing methods, including the beamformer-based search \cite{dmochowski2007generalized}, subspace methods \cite{schmidt1986multiple}, probabilistic generative mixture models \cite{rickard2002approximate}, and independent component analysis \cite{sawada2003direction}. 
However, it is hard to generalize these methods to \reviewaccepted{perform well under} real-world \reviewaccepted{conditions} with complex noise-reverberation interference and sources' spatial-temporal alternation.
Thanks to the advent of deep learning \cite{lecun2015deep}, one can further distill information inside the features which are extracted by these conventional methods. 
Since 2015, the number of DNN models for SSL is increasing explosively, covering all major types of network layer types, such as Multi-Layer Perceptrons (MLP) \cite{tsuzuki2013approach}, convolutional neural networks (CNN) \cite{hirvonen2015classification}, convolutional recurrent neural networks (CRNN) \cite{adavanne2019localization}, encoder-decoder neural networks \cite{le2019learning}, attention-based neural networks \cite{cao2021improved}, etc.
These state-of-the-art models will be \reviewaccepted{further} detailed in Section \ref{section_algorithm}.

\reviewaccepted{In many applications, these cutting-edge algorithms are required to be processed locally, for latency or privacy reasons.} 
\reviewaccepted{Such edge applications are} for example drone navigation \cite{hoshiba2017design}, hearing aids \cite{van2011sound}, and interactive robots \cite{trifa2007real}. 
These applications are expected to provide robust performance against harsh or varying environments during execution.  \reviewaccepted{Yet, these devices also suffer from} limited space available for the computational unit, \reviewaccepted{resulting in a need for a compute, memory, and energy-efficient design.}
\reviewaccepted{This requires a new class of SSL models, optimized} 
to run on a resource-constrained embedded device, yielding real-time outputs with limited \reviewaccepted{computational} performance, \reviewaccepted{memory} bandwidth, and power. 
Obviously, the mentioned requirements of SSL robustness and computation\reviewaccepted{al} efficiency create a tradeoff case, especially for \reviewaccepted{the typically compute-heavy} DNN models.

In terms of SSL robustness, \reviewaccepted{methods based on} steered response power \reviewaccepted{features} with phase transform filter (SRP-PHAT) \cite{dibiase2001robust} \reviewaccepted{lead the SotA} in SSL applications \reviewaccepted{in harsh environments}. Derived from the generalized cross-correlation (GCC) of microphone signals, the search among candidate locations with maximal SRP power is capable of tolerating noisy and reverberant environments. However, the original SRP-PHAT is computationally expensive \reviewaccepted{which makes it} impossible to meet the real-time requirements \reviewaccepted{in edge devices}. To relieve this, many modifications have been proposed to reduce SRP's complexity \cite{dietzen2020low}, enhance the parallelism \cite{minotto2013gpu}, optimize localization mechanism \cite{lima2014efficient} and etc. \reviewaccepted{Yet, computational requirements remain far above the capabilities of extreme edge devices.} 

\reviewaccepted{Moreover,} recently, the \reviewaccepted{combination} of Deep Neural Network (DNN) \reviewaccepted{models with the SRP features, results in cascaded} SRP-DNN models \cite{salvati2018exploiting, diaz2020robust} \reviewaccepted{that show further SSL accuracy and robustness improvements. Yet, this again comes at an increased computational burden.} 
Besides, SRP-PHAT sacrifices the spectral information of the source signal for its outstanding robustness. This further stresses the resource constraints, i.e. low-complexity demand, if auxiliary blocks are required to make up for such loss in complex missions. For instance, in sound event localization and detection (SELD) \cite{adavanne2018sound}, event classification DNN is jointly built beside localization to resolve overlapped multiple targets.

\reviewaccepted{As such, while providing excellent robustness and SSL performance,} 
the challenge of \reviewaccepted{bringing the} SRP-DNN method \reviewaccepted{for SSL to the} edge is twofold:
\begin{enumerate}
    \item The computation overhead: Both the SRP-map grid search and DNN inference \reviewaccepted{are}  time-consuming due to the computation amount and the data dependencies.
    \item The extremely-mixed acoustic scenes: Although SRP is designed to handle noisy and reverberant cases with robustness, it is challenging to make \reviewaccepted{compact, computational-efficient} DNN models converge on mixed cases with miscellaneous acoustic environments and randomly moving sources.
\end{enumerate}

In this paper, we \reviewaccepted{will start from} 
the Cross3D model \cite{diaz2020robust}, which is designed to robustly solve the challenge-(2) for single-source localization at indoor environments. To our knowledge, currently the Cross3D's dataset simulator \reviewaccepted{can} synthesize the widest range of indoor acoustic scenes, with randomness on noise levels, reverberation levels, source trajectories, sensor locations, room parameters and etc. On this dataset, Cross3D shows robust performance over random cases and outperforms other state-of-the-art models. However, the resulting Cross3D model becomes gigantic and fails for the challenge-(1), which will be elaborated in
Section \ref{section_methodologies} and \ref{section_experiments}. 

Therefore, we propose an optimized version of the Cross3D model towards edge-deployment hardware requirements. Firstly, we reveal the baseline Cross3D's bottlenecks in algorithm and computation. Secondly, we \reviewaccepted{assess and} exploit the model trade-off point between algorithmic accuracy and low-complexity computation. For the SRP part, we propose LC-SRP-Edge based on LC-SRP \cite{dietzen2020low} for lower hardware overhead. We then integrate this SRP-PHAT into the Cross3D model to replace the original lossy time-domain SRP. 
For the DNN part, we squeeze the original Cross3D to propose Cross3D-Edge, along with detailed ablation studies to discuss the impact on the model's robustness. \rebuttal{Thirdly, we discuss the hardware overhead and real-time processing capability of the proposed models with hardware modeling metrics, as well as physical edge device latency measurements. Finally, we provide a comprehensive comparison with other state-of-the-art research in the field of sound source localization. }

The rest of this paper is organized as follows. We revisit the details of related algorithms in Section \ref{section_algorithm}. We identify the baseline model's \reviewaccepted{computational} bottleneck and propose optimization methods in Section \ref{section_methodologies}. In Section \ref{section_experiments}, the proposed approach is evaluated against the baseline method on algorithm performance and hardware footprint, respectively. Then, in Section \ref{section_sota_comparison}, we compare our model with the state-of-the-art. Finally, Section \ref{section_conclusion} concludes the paper.

\begin{figure*}[t]
    \centering
    \setlength{\abovecaptionskip}{0.cm}
    \includegraphics[width=0.9\linewidth]{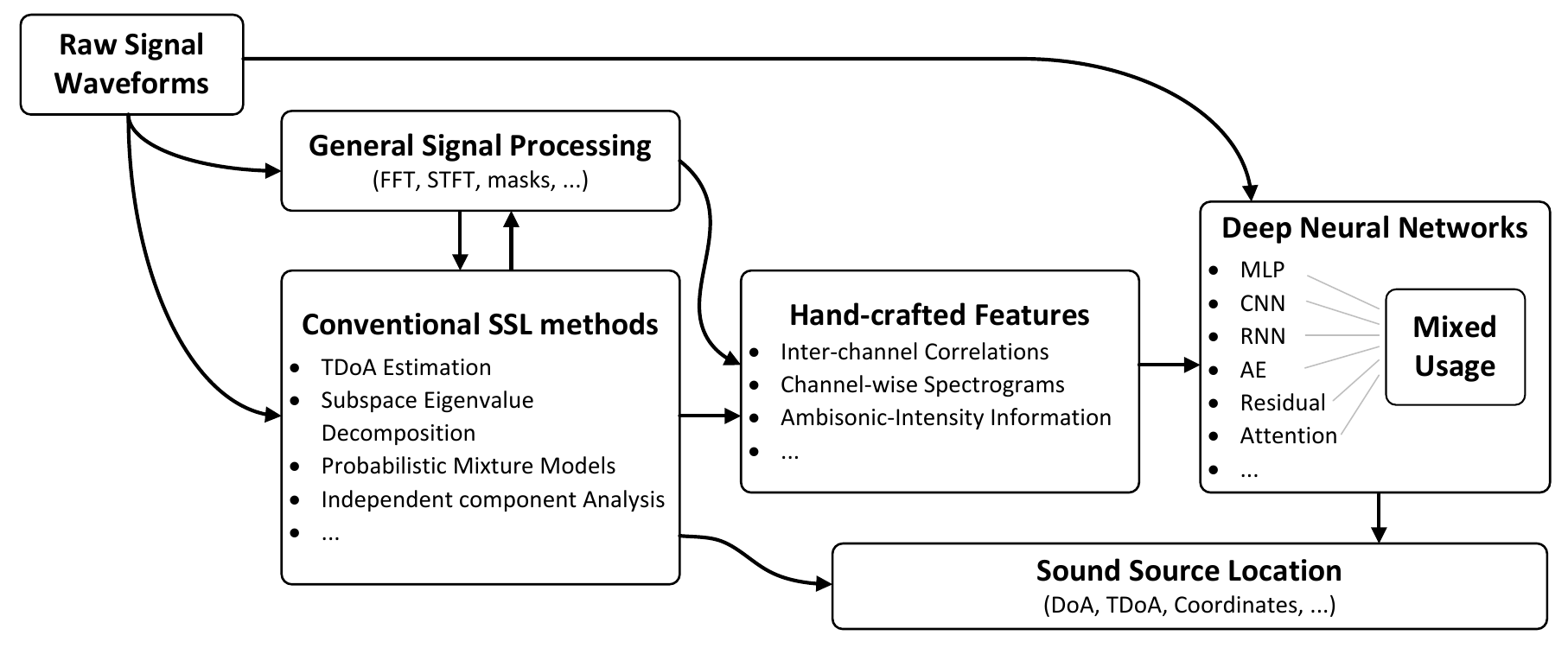}
    \caption{An overview diagram of the modern Sound Source Localization (SSL) practice with Deep Neural Networks (DNN). }
    \label{fig:SSL_DNN_intro}
\end{figure*}

\section{Related Algorithms}\label{section_algorithm}
In this section, we introduce the state-of-the-art algorithms in the region of SSL solutions \reviewaccepted{exploiting} DNN models. \reviewaccepted{The field is summarized in} 
Fig. \ref{fig:SSL_DNN_intro}. We start from the input features used by these SSL DNNs in Section \ref{section_algorithm_input_features}. Then we introduce different types of neural networks on how they contribute to the SSL solutions in Section \ref{section_algorithm_nn_types}. 
Finally, we describe the typical workflow of the SSL DNN system with reference to the Cross3D project \cite{diaz2020robust} in Section \ref{section_algorithm_system_overview}.

\subsection{Input Features}\label{section_algorithm_input_features}
As introduced in Section \ref{section_introduction}, most of the input features are derived from the baseline conventional SSL algorithms. Generally, the raw signal incorporates multi-channel audio sequences from a binaural or larger microphone array. Different input features \reviewaccepted{extracted from the raw audio} represent different aspects of raw signals \reviewaccepted{useful} for DNN reasoning. \reviewaccepted{Ordered by} increasing dependence on the reasoning power \reviewaccepted{of the DNN}, three directions can be categorized: the inter-channel relationships, the channel-wise spectrograms, the original acoustic features.

The first direction is to \reviewaccepted{extract features characterizing} the inter-channel relationships and differences. Based on the different spatial positions of each \reviewaccepted{microphone} sensor, one can study the signal channels in pairs and infer the source location from indirect metrics such as the time difference of arrival (TDoA) \cite{xu2012high} peak searching. 
Based on the TDoA, the generalized cross-correlation with phase transform (GCC-PHAT) \cite{knapp1976generalized} is one of the mostly-used features in the search. Furthermore, the steered response power with phase transform (SRP-PHAT) \cite{dmochowski2007generalized} is designed to have better tolerance of noise and reverberation, as SRP-PHAT measures the ``energy'' across the entire microphone array instead of microphone pairs in GCC-PHAT. 
They triggered the most famous traditional methods like MUSIC \cite{schmidt1986multiple} and ESPRIT \cite{roy1989esprit} at the beginning of SSL research. Afterwards, a lot of DNN-based methods follow \cite{xiao2015learning, li2018online, noh2019three, comanducci2020source, diaz2020robust}, \reviewaccepted{demonstrating that} 
the interaural difference features of binaural signals are also useful representations for DNN inference \cite{youssef2013learning, roden2015sound, sivasankaran2018keyword}.

The second direction is channel-wise feature processing. Leaving the inter-channel characteristics for DNN reasoning, these features focus on spectra and temporal information. As a result, short-term Fourier transformation (STFT) is commonly used on individual signal channels with consecutive frames \cite{vincent2018audio}. As a spectrogram-based feature family, different aspects of the feature are proved to be useful for DNN-based SSL systems, including magnitude spectrograms \cite{yalta2017sound, wang2018robust}, phase spectrograms \cite{subramanian2021deep,zhang2019robust}, Mel-scale spectrograms \cite{vecchiotti2018deep, kong2019cross}, and the concatenation of these \cite{schymura2021pilot, guirguis2021seld}. 

The third direction is the original acoustic features of sound. 
On the one hand, the Ambisonic representation format \cite{jarrett2017theory} directly contains the spatial information of a sound field. 
That means the SSL system no longer needs to use the sensor array configuration to reconstruct this field. In practice, first-order Ambisonics (FOA) \cite{kapka2019sound, jee2019sound, adavanne2018direction} and higher-order Ambisonics (HOA) \cite{varanasi2020deep, poschadel2021direction} are used in neural-based algorithms. 
On the other hand, the sound intensity feature is capable of depicting the gradient of the phase of sound pressure, which is usually used together with the Ambisonics features \cite{perotin2018crnn, yasuda2020sound}. 
Moreover, some recent research directly feeds raw signal waveforms to the DNN model \cite{pujol2019source, huang2020time, pujol2021beamlearning}, expecting the DNN to rule out better features than hand-crafted ones.  

In this paper, we choose the SRP-PHAT approach as it is well studied and dedicated for robust SSL missions, such that a good baseline to begin hardware optimization for extreme edge platforms. Its robustness is proved by Cross3D \cite{diaz2020robust} on harsh and extremely mixed acoustic environments. Other mentioned features, keeping more detailed spatial or spectral information, usually lead to accuracy compromises between sound localization and classification, or bring about greater algorithm complexity than SRP-PHAT to generalize various acoustic scenes. 

\subsection{Neural Network Types}\label{section_algorithm_nn_types}
\reviewaccepted{Modern SSL solutions feed the features discussed in the previous subsection into a trained neural network.} Similar to the input features, multiple types of neural network layers are employed to build models for SSL problems. A comprehensive survey is available at \cite{grumiaux2021survey}.  

The \reviewaccepted{initial} type of DNN models, Multiple Layer Perceptrons (MLP), was used in the early stage of solving SSL problems with deep learning \cite{kim2011direction, tsuzuki2013approach, youssef2013learning}. Since the convolutional neural network (CNN) showed its power in pattern recognition, CNN-based algorithms have been proposed to extract hidden SSL information and rule out DOA estimations from almost every input feature in Section \ref{section_algorithm_input_features}, such as the magnitude spectrograms \cite{hirvonen2015classification}, phase spectrograms \cite{chakrabarty2017multi, chakrabarty2017broadband}, binaural features \cite{thuillier2018spatial}, raw input signals \cite{vera2018towards}, GCC-PHAT \cite{varzandeh2020exploiting}, SRP-PHAT \cite{pertila2017robust, diaz2020robust}, etc.
They prove neural networks' capability to surpass the conventional methods in SSL. Later, recurrent layers have been applied, including long short-term memory (LSTM) \cite{hochreiter1997long} and gated recurrent units (GRU) \cite{cho2014learning} \reviewaccepted{to incorporate timing information through states}. For SSL problems, convolutional recurrent neural networks (CRNN) are widely used instead \cite{adavanne2018direction, adavanne2018sound, kapka2019sound}, in order to extract spatial-temporal information at the same time.

Based on the convolutional and recurrent DNN layers, other deep learning techniques are incorporated to increase system accuracy. On the one hand, the residual connections are added to \reviewaccepted{improve training convergence on} deeper neural networks.
\cite{suvorov2018deep, naranjo2020sound} show how residual CNN models outperform the conventional in SSL. \cite{shimada2020sound, shimada2021accdoa, wang2020ustc} comprehensively use CRNN layers with residual connections and succeed in complicated missions such as the SELD in the DCASE2020 challenge.
On the other hand, attention-based mechanisms also benefit the SSL field for their capability of understanding the acoustic environmental context. 
For example, \cite{cao2021improved, schymura2021pilot, wang2021four} uses multi-head self-attention layers with the Transformer architecture \cite{vaswani2017attention} to detect and estimate the source location when multiple sound events are mixed together. Finally, encoder-decoder neural networks (AE) \reviewaccepted{have proven beneficial} because of their unsupervised learning capabilities in cases with little knowledge about the sound source. As generative models, AE-based methods \cite{le2019learning, comanducci2020source, wu2021sslide} solve SSL problems by separating the sound features to each candidate region. 

However, with more and more feature types and neural network layers fused in the latest models, the computational efficiency and model parallelism drop drastically. For instance, in \cite{guirguis2021seld}, the authors focused on improving the hardware friendliness of SELDnet \cite{adavanne2018sound} by replacing the recurrent blocks with temporal convolutional network (TCN) layers. The resulting SELD-TCN is proved to yield the same-level accuracy against the baseline while greatly improving the latency.
In this paper, we also aim at hardware overhead reduction without giving in to model robustness and accuracy. We choose the Cross3D \cite{diaz2020robust} as our baseline for two reasons: 1) It is proved to maintain robust performance across harsh and varying acoustic scenes. 2) It is a fully-causal CNN model which is hardware-friendly such that forms a good baseline in terms of real-time processing on edge. 
The detailed analysis and comparison of the hardware efficiency lie in Section \ref{section_experiments_hardware}.

\subsection{System Overview}\label{section_algorithm_system_overview}
Here we describe the workflow of a typical SSL DNN system under the Cross3D project \cite{diaz2020cross3dproject}. Similar to other DNN frameworks, the workflow consists of dataset preparation, input feature calculation, neural network training-testing module, and supporting pre/post-processing modules. In the field of SSL, both synthesized and real-world datasets are considered, such as the dataset series in the DCASE2020 challenge Task3 \cite{politis2020dataset}. While recorded datasets include realistic and rich environmental features, synthesized datasets provide wider coverage of recording cases under certain acoustic scenes.

\begin{figure*}[t]
    \centering
    \includegraphics[width=0.8\linewidth]{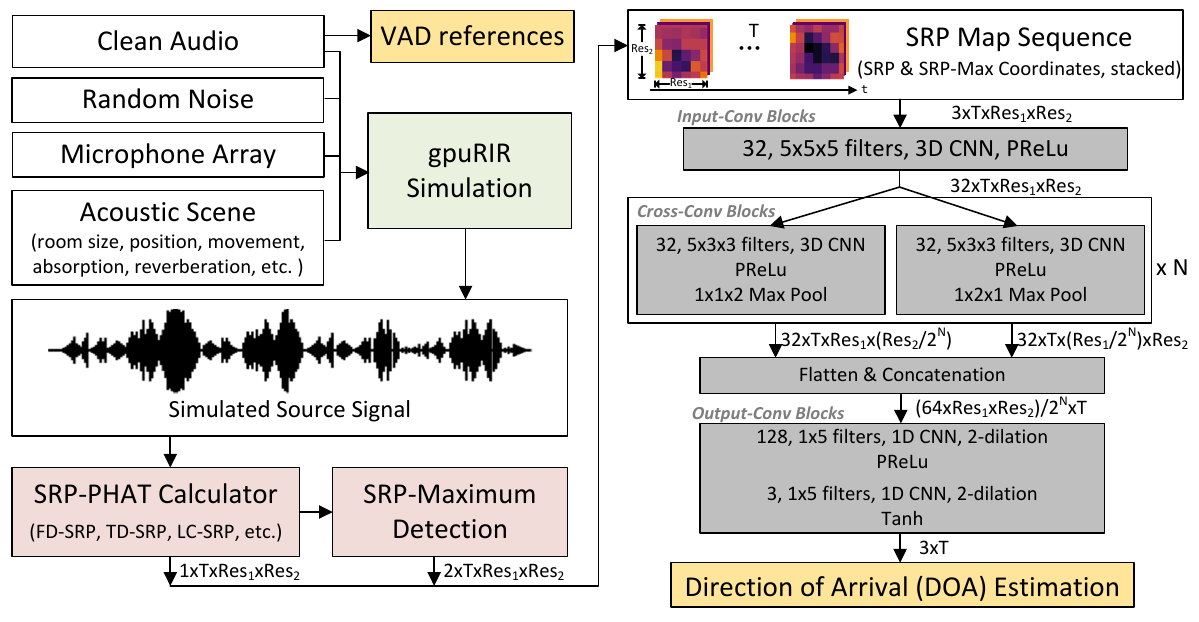}
    \setlength{\abovecaptionskip}{0.cm}
    \caption{Cross3D \cite{diaz2020robust} model structure and workflow. $T$ denotes the length of SRP sequence. The branch depth $N$ is determined by SRP resolution $N = \min(4, \log_2(\min(Res1, Res2))$. 
    }
    \label{fig:cross3d-workflow}
\end{figure*}

As shown in Fig. \ref{fig:cross3d-workflow}, the Cross3D workflow focuses on a synthesized indoor dataset from a GPU-based simulator named gpuRIR \cite{diaz2021gpurir}. During simulation, clean audio files (dry signal dataset) are fetched with random noise signals to form the original source signal. Then, a runtime-generated environment configuration is attached to the source signal, such as the room size, source-sensor position\&movement, noise-reverberation ratio, room surface absorption, etc. Afterwards, the simulator (gpuRIR) will generate the room impulse response (RIR) based on the image source method (ISM) \cite{allen1979image} and the microphone array topology. Finally, the original source signals are convoluted with the RIRs to build multi-channel microphone recordings, serving as the input to SSL problems. 

With input signals ready, corresponding DNN features are calculated, such as the SRP-PHAT and its maximums in Fig. \ref{fig:cross3d-workflow}. The Cross3D's feature map is built by stacking the SRP-PHAT feature map and its maximum coordinates into a 5D tensor. 
Further, the training and testing of DNN models are triggered in a pipelined manner.

\reviewaccepted{It is important to note that} it is common for SSL systems to involve peripheral pre/post-processing modules. For example, the dry signal dataset in Cross3D is Librispeech \cite{panayotov2015librispeech} which contains intervals between human voice audios. Hence, a voice activity detection (VAD) module is implemented to mark the active sound frames. The VAD reference indices are taken into account when testing the accuracy of SSL on sequential sound snippets. Besides, other modules can also be added in pre-processing, such as sound source separation when applying single-source SSL models to multiple-source datasets. 

In this paper, we focus on the optimization and discussion of the accuracy-efficiency trade-off for the Cross3D structure. Hence, the later experiments follow the workflow of the original project (Fig. \ref{fig:cross3d-workflow}).

\section{Methodologies}\label{section_methodologies}
For edge deployment, SSL applications must satisfy the device resource constraints and real-time execution requirements. 
Hence, we first review the details of the original Cross3D \cite{diaz2020robust} in Section \ref{section_methodologies_original_cross3D}, including Section \ref{section_methodologies_original_cross3D_input} the SRP-PHAT input features, Section \ref{section_methodologies_original_cross3D_nn} the neural network structure, and Section \ref{section_methodologies_original_cross3D_bottleneck} the bottleneck identification.
Based on these, we further propose our \textbf{LC-SRP-Edge} and \textbf{Cross3D-Edge} for the input feature computation and neural network structure in Section \ref{section_methodologies_ours}.
Finally, we summarize the complexity of these algorithms and deduce the related hardware overhead in Section \ref{section_methodologies_overhead}.

\subsection{\reviewaccepted{Assessing the o}riginal Cross3D Model}\label{section_methodologies_original_cross3D}
\begin{figure*}[t]
    \centering
    \includegraphics[width=0.8\linewidth]{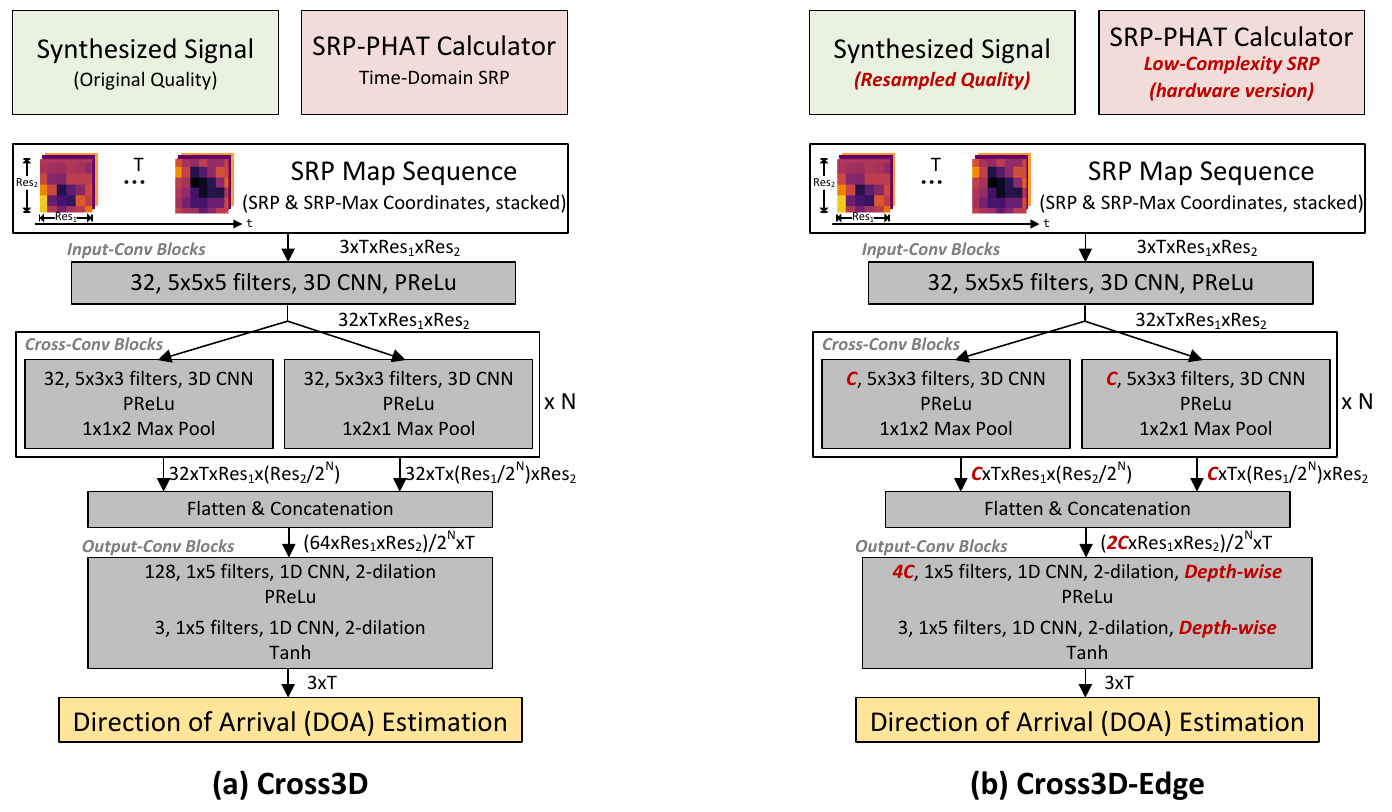}
    \setlength{\abovecaptionskip}{0.cm}
    \caption{Diagrams of the original \textbf{Cross3D baseline} model (a) and the proposed \textbf{Cross3D-Edge} model (b). Res1 and Res2 denotes the SRP's candidate space resolution on the dimension of elevation and azimuth, respectively. The modifications of the algorithm are marked in red text. 
    }
    \label{fig:cross3d-methodologies}
\end{figure*}

The overview of the \reviewaccepted{baseline} Cross3D model is shown in Fig. \ref{fig:cross3d-methodologies} (a).

\subsubsection{Input Feature}\ \label{section_methodologies_original_cross3D_input}

As introduced in Section \ref{section_algorithm}, the Cross3D \reviewaccepted{DNN} consumes an SRP-PHAT feature map as the input representation of microphone signals. With the dry clean audio source from Librispeech, sampled at 16kHz and synthesized to a 12-microphone array, Cross3D computes the spectral features via the real-value Fourier transform on a 4096-sample 25\%-overlap Hanning window. 
After that, the SRP-PHAT map is obtained via the temporal-domain SRP algorithm (TD-SRP) in the original Cross3D project.

The central idea of SRP is to compute the power output of a filter-and-sum beamformer that virtually steers the microphone array towards candidate positions. The original SRP-PHAT \cite{dibiase2001robust} is obtained from frequency-domain operations on the Fourier transformed signal $X(\omega)$. Considering microphone pairs $(m, m^{\prime})$ within a $M$-microphone array, the SRP-PHAT $\mathcal{P}(q)$ can be given as
\begin{equation}
    \mathcal{P}(\textbf{q}) = \sum_{(m, m^{\prime}):m > m'} \int \frac{X_m(\omega)X_{m^{\prime}}^*(\omega)}{\left| X_m(\omega)X_{m^{\prime}}^*(\omega) \right|}e^{j\omega(\tau_m^\mathrm{\textbf{q}}-\tau_{m^{\prime}}^\mathrm{\textbf{q}})}d\omega\\
\label{eq:srp_phat_origin}
\end{equation}
where \textbf{q} belongs to the candidate location set $\mathbb{Q}$ and $(\tau_m^\mathrm{\textbf{q}}-\tau_{m^{\prime}}^\mathrm{\textbf{q}})$ represents the time difference of arrival (TDOA) of pair $(m, m^{\prime})$ on source location \textbf{q}. Further, with the definition of generalized cross correlation (GCC), we can rewrite Eq. (\ref{eq:srp_phat_origin}) with frequency-domain GCC-PHAT $\mathcal{G}_{m, m^{\prime}}(\omega)$ as
\begin{equation}
    \mathcal{P}(\textbf{q}) = \sum_{(m, m^{\prime}):m > m'} \int \mathcal{G}_{m, m^{\prime}}(\omega)e^{j\omega(\tau_m^\mathrm{\textbf{q}}-\tau_{m^{\prime}}^\mathrm{\textbf{q}})}d\omega
\label{eq:srp_phat_origin_gcc}
\end{equation}

To eliminate the huge computation of integrating over all frequency bins, one can first perform inverse Fourier transformation of $\mathcal{G}_{m, m^{\prime}}(\omega)$ to calculate SRP in the time domain (TD) as  
\begin{equation}
    \mathcal{P}(\textbf{q}) = 2 \sum_{(m, m^{\prime}):m > m'} \mathcal{G}_{m, m^{\prime}}(\Delta t_{m, m^{\prime}} (\textbf{q}))
\label{eq:srp_phat_td}
\end{equation}
where $\mathcal{P}(\textbf{q})$ is the $\textbf{q}_{th}$  TD-SRP, $\mathcal{G}_{m, m^{\prime}}$ is the corresponding TD-GCC and $\Delta t_{m, m^{\prime}} (\textbf{q})$ is the indexing term from $(\tau_m^\mathrm{\textbf{q}}-\tau_{m^{\prime}}^\mathrm{\textbf{q}})$. 

TD-SRP could reduce the complexity from the level of discrete Fourier transform (DFT) in Eq. (\ref{eq:srp_phat_origin}) to the fast Fourier transform. It is therefore adopted in many SRP-based algorithms, such as this Cross3D project \cite{diaz2020cross3dproject}.

\subsubsection{Neural Network Structure}\ \label{section_methodologies_original_cross3D_nn}

To extract both the spatial and temporal features of a moving sound source, Cross3D uses causal convolution layers with 1 kernel axis for the time dimension. Shown in Fig. \ref{fig:cross3d-methodologies} (a), we can denote the network with 3 major blocks: \textit{Input\_Conv}, \textit{Cross\_Conv}, and \textit{Output\_Conv}. 

In the \textit{Input\_Conv} block, the input is \reviewaccepted{a stack of $T$ consecutive audio frames, where each frame consists of the time step's} SRP map \reviewaccepted{with resolution $Res_1\times Res_2$} together with the maximum's coordinates of the map (2D normalized coordinates, forming the other 2 channels of the input feature).
This input SRP \reviewaccepted{tensor} is processed by a 3D CNN layer with 32 filters of size 5$\times$5$\times$5. 
The activation function used in Cross3D is PReLu \cite{he2015delving}.

Following the \textit{Input\_Conv}, the \textit{Cross\_Conv} block is formed by several consecutive 3D CNN layers in 2 parallel branches. Each layer incorporates 32 filters of size 5$\times$3$\times$3, the PReLu activation, and a max-pooling layer. The difference between the two branches is the \reviewaccepted{direction} of max pooling, which is 1$\times$1$\times$2 and 1$\times$2$\times$1, respectively. This forces the network to extract higher-level SRP features on both the azimuth and elevation dimensions separately. The amount of each branch's stacked CNN layers is defined by  $N = \min(4, \log_2(\min(Res1, Res2))$ to avoid computation error.  

After the \textit{Cross\_Conv} operations, the final \textit{Output\_Conv} block finishes the inference. In this block, 1D CNNs are invoked for the temporal feature aggregation, with filters of size 1$\times$5 and dilations of size 2. Hence, \reviewaccepted{to enable this} \textit{Cross\_Conv} outputs are flattened and concatenated into 1-dimensional temporal features \reviewaccepted{before being fed to the  \textit{Output\_Conv} block}. Finally, the DOA estimation is generated as 3D Cartesian coordinates ($T$-frame xyz). 

\subsubsection{Bottlenecks}\ \label{section_methodologies_original_cross3D_bottleneck}

The basic idea of Cross3D is to train one single model that supports a wide variety of acoustic environments and sound sources.
Proved in the original literature \cite{diaz2020robust}, Cross3D's practice to combine SRP-map features with a DNN back-end brings increased robustness, yet at the cost of system complexity. 
From the diagram in Fig. \ref{fig:cross3d-methodologies}, the reader can have an intuition about the bulk size of the Cross3D network. Moreover, not only does the higher $Res1\times Res2$ feature map invoke a huge computation load, but Cross3D's accuracy no longer improves in such cases as well.

\reviewaccepted{When we assess}
the original Cross3D's SSL accuracy as shown in Table \ref{tab:original_cross3d_efficiency}, we can identify the two bottlenecks mentioned above: 

\begin{table}[t]
\setlength{\abovecaptionskip}{0.cm}
\caption{Comparison on the original Cross3D's SSL Efficiency.}
\vspace{-0.cm}
\label{tab:original_cross3d_efficiency}
\includegraphics[width=\linewidth]{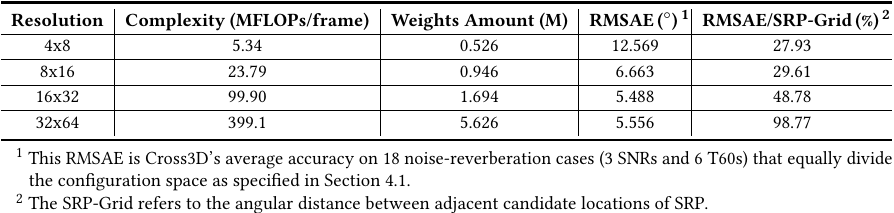}
\end{table}

The first bottleneck is the accuracy saturation at higher resolutions. 
Although we can achieve or retain low SSL errors when selecting higher-resolution SRP maps (E.g. the 32$\times$64) for the model, \reviewaccepted{it does not outperform medium-resolution solutions}.
For a multi-resolution model, we take the angular distance between adjacent SRP candidate locations (the ``SRP-Grid'') as a reference threshold. 
Highlighted in Table \ref{tab:original_cross3d_efficiency}'s ``RMSAE/SRP-Grid'' column, the considerable increase of this metric means that the Cross3D fails to recognize the differences between finer-grained adjacent locations.

The second bottleneck is the network size explosion. As shown in Table \ref{tab:original_cross3d_efficiency}, the size of Cross3D increases rapidly at higher-resolution cases, in both the computational complexity and weight amount aspects.
Along with the accuracy saturation, these extra overheads are actually wasteful.

\subsection{Proposed Methods}\label{section_methodologies_ours}
Based on the analysis of the original Cross3D in Section \ref{section_methodologies_original_cross3D}, we propose the \textbf{LC-SRP-Edge} (Section \ref{section_methodologies_ours_input}) and \textbf{Cross3D-Edge} (Section \ref{section_methodologies_ours_nn}) towards better accuracy and more efficient computation.  

\subsubsection{SRP-PHAT Complexity}\label{section_methodologies_ours_preface}\

Although the original Cross3D's TD-SRP Eq. (\ref{eq:srp_phat_td}) is much simpler to compute than FD-SRP (Eq. (\ref{eq:srp_phat_origin})) with techniques like the fast Fourier transformation (FFT), the quantization of TDOAs to get integer indices $\Delta t_{m, m^{\prime}} (\textbf{q})$ makes TD-SRP mathematically lossy \cite{minotto2013gpu}. Hence, one can refer to interpolation methods \cite{do2007real,tervo2008interpolation} to reduce such loss. However, this would in return lead to additional computations.

In this paper, we will replace this SRP calculation with a low-complexity SRP (\textbf{LC-SRP}) \cite{dietzen2020low}, which uses the Whittaker-Shannon interpolation  \cite{marks2012introduction} on TD-GCC elements $\mathcal{G}_{m, m^{\prime}}$ for a perfect reconstruction of Eq. (\ref{eq:srp_phat_origin}). Assuming the microphone signal is bandlimited by $\omega_0$ \cite{dietzen2020low}, this approximation can be calculated as 
\begin{align}
    \mathcal{G}_{m, m^{\prime}}^{appr}(\tau)&= \sum_{n\in\mathbb{N}_{m,m^{\prime}}} \mathcal{G}_{m, m^{\prime}}(nT) sinc(\tau/T-n)\label{eq:lc_srp_1}\\
    &= \sum_{n\in\mathbb{N}_{m,m^{\prime}}} \sum_{k=0}^{K-1} 2 \Re \left[\mathcal{G}_{m, m^{\prime}}(k) e^{j\frac{2\pi k}{K}nT}\right]sinc(\tau/T-n) \label{eq:lc_srp_2}
\end{align}
with $T=2\pi/\omega_0$ the critical sampling period, $K$ the number of frequency bins, $\tau$ the target TDOAs, and $\mathbb{N}_{m,m^{\prime}}$ the number of sampled frequency bins for the $(m,m^{\prime})$-th microphone pair. 
One can notice that $\mathbb{N}_{m,m^{\prime}}$ differs between microphone pairs. Given a microphone pair $m,m^{\prime}$ with the pair distance of $Dist_{m,m^{\prime}}$, audio sampling rate $fs$ and speed of sound $c$, the sample index $n$ satisfies  $n \in [-N_{samp}(m,m^{\prime}),\, N_{samp}(m,m^{\prime})], n\in\mathbb{Z}$, where
\begin{equation}
    N_{samp}(m,m^{\prime})=\lfloor\frac{Dist_{m,m^{\prime}}}{c}\cdot fs\rfloor \label{eq:lc_srp_nsamp}
\end{equation}

Now we can estimate the computational complexity to calculate one SRP-PHAT feature map with TD-SRP and LC-SRP from Eqs. (\ref{eq:srp_phat_origin_gcc}), (\ref{eq:srp_phat_td}) and (\ref{eq:lc_srp_2}). Among the following estimation, both the Fourier and sinc coefficients are pre-computed and reused. 

Let us assume an SRP application case with $N$ microphones, $K$ signal Fourier transformation points, $Q$ SRP candidate positions, and $N_{samp}$ LC-SRP interpolation indices. As a result, we have $P=\frac{N(N-1)}{2}$ microphone pairs and $(\frac{K}{2}+1)$ frequency bins for real-valued source signals. Then the common computational complexity among these methods, which is to calculate frequency-domain GCC-PHATs (Eq. (\ref{eq:srp_phat_origin_gcc})), can be denoted as: 
\begin{enumerate}
    \item Real signal FFT: \quad$2N\cdot Klog_2K$
    \item Frequency-domain GCC: \quad$4\cdot\frac{N(N-1)}{2}\cdot (\frac{K}{2}+1)$
    \item PHAT normalization: \quad$10N\cdot (\frac{K}{2}+1)$
\end{enumerate}
Note that in this paper, we count 1 real-valued Multiply-Accumulate (MAC) as 2 arithmetic operations (OPs).
Then for the TD-SRP in Eq. (\ref{eq:srp_phat_td}), the further computation is formed by the reduction operation over inverse Fourier transformed GCC-PHATs, 
\begin{enumerate}
    \item GCC-PHAT IRFFT: \quad$2\cdot\frac{N(N-1)}{2}\cdot Klog_2K$
    \item TD-SRP: \quad$\frac{N(N-1)}{2}\cdot Q$
\end{enumerate}
While in LC-SRP's definition Eq. (\ref{eq:lc_srp_2}), the most expensive computation is the frequency-domain inverse discrete Fourier transformation:
\begin{enumerate}
    \item GCC-PHAT IDFT: \quad $N_{samp}\cdot (2K+4)$
    \item Sinc interpolation: \quad $N_{samp}\cdot (2Q)$
\end{enumerate}
$N_{samp}=\sum \mathbb{N}_{m,m^{\prime}}$ is the total number of sampled frequency bins for the entire sinc interpolation. Typically, we can find $\frac{N(N-1)}{2} < N_{samp} \leq \frac{N(N-1)}{2}\cdot(\frac{K}{2}+1)$. Hence, we can notice that the LC-SRP's computation is more efficient at lower SRP resolution ($Q$) and compact microphone arrays (i.e. small $N_{samp}(m,m^{\prime})$ for each pair).

\subsubsection{LC-SRP-Edge}\label{section_methodologies_ours_input}\ 

In spite of LC-SRP's complexity reduction, it requires additional memory cost.
According to Eqs. (\ref{eq:lc_srp_1}) and (\ref{eq:lc_srp_2}), the sinc coefficients \bm{$sinc(\tau/T-n)$} are aperiodic across various aspects, including the dimensions of microphone pairs, sampling points, and SRP candidates. Hence, for one SRP map, the sinc coefficient amount we need is:
\begin{equation}
    Sinc\_Amount = max(N_{samp}(m,m^{\prime}))\cdot\frac{N(N-1)}{2}\cdot Q
\end{equation}
This would result in a large memory overhead.
For example, this overhead is 0.84 MByte for 32bit sinc coefficients, if using an 10-microphone array with maximal pair distance of 0.1 meter, recording audio under 16kHz, and computing a 1000-dot SRP map.

Therefore, we propose \textbf{LC-SRP-Edge} to efficiently boost LC-SRP implementation for edge hardware. Inspired by Eq. (\ref{eq:lc_srp_2}), we further optimize the complexity of LC-SRP by pairing the interpolations. Considering the 0-symmetric interpolation indices from Eq. (\ref{eq:lc_srp_nsamp}) and the complex-conjugate nature of Fourier transformation coefficients, we expand and rewrite Eq. (\ref{eq:lc_srp_2}) as:
\begin{equation}
\begin{aligned}
    \mathcal{G}_{m, m^{\prime}}^{appr}(\tau) =    \sum_{n=0}^{N_{samp}(m,m^{\prime})} &\left\{
    2\frac{sin(\pi\,\tau/T)\odot (\tau/T)}{\pi (\tau/T-n)(\tau/T+n)} \cdot \sum_{k=0}^{K-1} \Re \left(\mathcal{G}_{m, m^{\prime}}(k)\right) \cdot \Re \left(e^{j\frac{2\pi k}{K}nT}\right) \right.\\
    &\left.+ 2\frac{sin(\pi\,\tau/T)\odot n}{\pi (\tau/T-n)(\tau/T+n)} \cdot \sum_{k=0}^{K-1} \Im \left(\mathcal{G}_{m, m^{\prime}}(k)\right) \cdot \Im \left(e^{j\frac{2\pi k}{K}nT}\right) \right\} \cdot cos(n\pi)\label{eq:lc_srp_new}
\end{aligned}
\end{equation}
where $\odot$ denotes the element-wise product and $\Re/\Im$ denotes the real/imaginary part, respectively. As a result, we can extract the common factor in Eq. (\ref{eq:lc_srp_new}) as the new pre-computed interpolation coefficients:
\begin{equation}
    \textbf{W}_{sinc}^n = 2\frac{sin(\pi\,\tau/T)}{\pi (\tau/T-n)(\tau/T+n)}, \quad\quad n\in[0, N_{samp}(m,m^{\prime})], n\in \mathbb{Z}\label{eq:lc_srp_new_weight}
\end{equation}

By reducing the indexing range of $n$ from $[-N_{samp}(m,m^{\prime}),\, N_{samp}(m,m^{\prime})]$ to $[0, N_{samp}(m,m^{\prime})]$, Eqs. (\ref{eq:lc_srp_new}) and (\ref{eq:lc_srp_new_weight}) would reduce approximately 50\% the computation and memory space for LC-SRP's interpolation. 
Mathematically equivalent to the original LC-SRP in Eq. (\ref{eq:lc_srp_2}), this upgraded equation  can reduce the computational complexity from $N_{samp}\cdot (2K+4+2Q)$ to:
\begin{equation}
\begin{aligned}
    Complexity_{LC-SRP-Edge} &= Complexity(n=0) + Complexity(n \neq 0)\\ 
    &= \frac{N(N-1)}{2}\cdot(\frac{K}{2}+1+Q) + (N_{samp}-\frac{N(N-1)}{2})\cdot\frac{2K+4+4Q}{2}\\
    &= (N_{samp}-\frac{N(N-1)}{4})\cdot(K+2+2Q) \label{eq:complexity-lcsrp-edge}
\end{aligned} 
\end{equation}
Especially for $Complexity(n=0)$, the imaginary part is also discarded as $\Im [exp(j\frac{2\pi k}{K}nT)]\equiv 0$.
Please note that further reuse of sinc coefficients could be possible if considering symmetrical structures in the microphone array topology. However, this falls into the dedicated optimization which is beyond the scope of this paper.

Last but not least, one can also notice that the quality of input signal, \textit{i.e.} the signal sampling rate $fs$ and windowed points $K$, is also a dominant factor regarding the above complexity equations. However, different from the mathematically equivalent optimization above, the downsampled audio would result in lossy SRP maps again. Hence, the tradeoff between SSL accuracy and complexity when reducing the $K-fs$ factor group will be evaluated in Section \ref{section_experiments_algorithm_param} and Section \ref{section_experiments_hardware}
, along with more detailed comparisons of TD-SRP, LC-SRP, and LC-SRP-Edge. 

\subsubsection{Cross3D-Edge}\label{section_methodologies_ours_nn}\ 

In Section \ref{section_methodologies_original_cross3D_bottleneck}, we show the bottleneck in the original Cross3D model, which is the SSL accuracy saturation and network size explosion at higher resolutions. 
From Table \ref{tab:original_cross3d_efficiency}, one can notice that the \textbf{8x16} resolution case turns out to be a good tradeoff point. Hence, we start with this finding for further optimizations.

\begin{figure*}[t]
    \centering
    \includegraphics[width=0.9\linewidth]{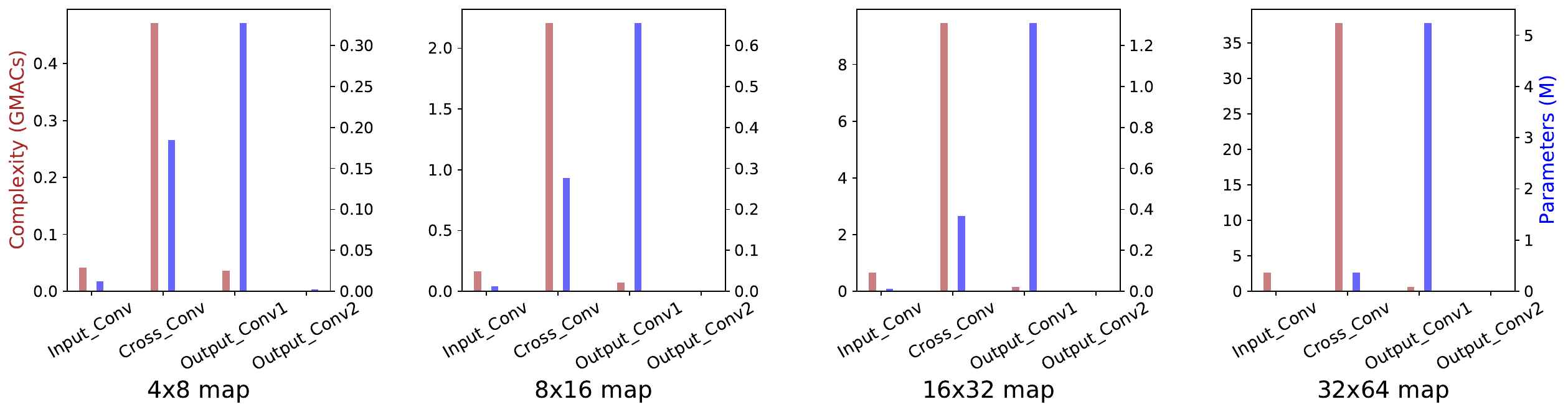}
    \setlength{\abovecaptionskip}{0.cm}
    \caption{The computational-complexity and parameter-amount distributions of the original Cross3D \cite{diaz2020robust} across network layers, demonstrating the fact that \textit{Cross\_Conv} is the most computationally-intensive while \textit{Output\_Conv1} is the most memory-expensive. Note that the layer name is in line with the diagram in Fig. \ref{fig:cross3d-methodologies}, where \textit{Output\_Conv1} and \textit{Output\_Conv2} stands for the last two 1D CNN layers, respectively.
    }
    \label{fig:workload-distribution-original}
\end{figure*}

We first profile the network structure to break down the composition of Cross3D's workload overhead.
Shown in Fig. \ref{fig:workload-distribution-original}, computation complexity and storage overhead for weight parameters comes from the \textit{Cross\_Conv} and \textit{Output\_Conv1} layers, respectively. 
Resulting from the consecutive 3D CNN layers in \textit{Cross\_Conv} and huge input channel size in \textit{Output\_Conv1}, this phenomenon is even more obvious at higher resolutions.
Therefore, we see the potential and necessity of modifying the Cross3D topology for these bottlenecks. 

On the one hand, we intend to squeeze the model along the output-channel dimension of several layers denoted as ``\bm{$C$}'' in Fig. \ref{fig:cross3d-methodologies} (b) to reduce the $O(C^2)$ complexity of the \textit{Cross\_Conv} layers. 
At the same time, we change the output channel size of \textit{Output\_Conv1} to $4C$. In order to stay in line with the shape of original Cross3D, we keep the ratio between the output channel sizes of \textit{Cross\_Conv} and \textit{Output\_Conv1}.

On the other hand, to further reduce the memory overhead for the weights in \textit{Output\_Conv1}, we adopt the depth-wise separable convolution \cite{chollet2017xception} to replace the original 1D CNN. Originally, the \textit{Output\_Conv1} layer has an input channel size >1000 and an output channel size >100.
Hence, based on the nature of depth-wise separable convolution, huge amount of weights would be saved after this modification.

We name this optimized model as \textbf{\textit{Cross3D-Edge}}. In the next sections, we conduct ablation experiments to elaborate on how our optimizations influence the model performance and achieve better algorithm-hardware trade-offs.
Considering the fact that acoustic environments change from time to time, it is also interesting to work out an adaptive model structure that efficiently handles the varying scenes. 

\begin{table}[t]
\setlength{\abovecaptionskip}{0.cm}
\caption{The summary of SRP computational complexity and parameter amount to be cached at on-chip memory of the three algorithms described in Section \ref{section_methodologies_original_cross3D_input} and \ref{section_methodologies_ours_input}.}
\vspace{-0.cm}
\label{tab:hardware-overhead-summary}
\includegraphics[width=\linewidth]{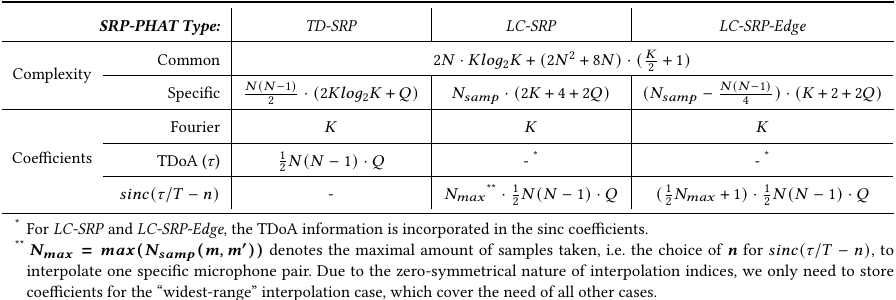}
\end{table}

\subsection{Hardware Overhead}\label{section_methodologies_overhead}
To assess the consequences of the proposed modifications in terms of hardware efficiency, we summarize the computational complexity and coefficient volume of the three SRP-PHAT algorithms in Table \ref{tab:hardware-overhead-summary}. Besides, the hardware footprint of the succeeding neural network back-end, in terms of the number of network weights and operations, can be obtained through the deep learning framework profiling tools, such as the PyTorch profiler. The detailed data for Cross3D and Cross3D-Edge versions are reported in Section \ref{section_experiments}.

For hardware evaluation purposes, we characterize the memory footprint in terms of necessary on-chip memory space to undertake these algorithms for real-time execution. \reviewaccepted{To streamline and align the estimation of the different algorithmic alternatives}, we assume an ideal hardware mapping strategy: 
\begin{enumerate}
    \item \textit{The SRP part}: For each SRP-PHAT map, the multi-channel input signal is updated and fetched from the main memory at the start of every windowed Fourier transform. The on-chip memory overhead includes the input-output data, intermediate variables, and the SRP-specific coefficients. We assume the resulting SRP-PHAT is directly consumed by the DNN computation unit.
    \item \textit{The DNN part}: Based on the nature of causal convolution, information of certain past frames is needed for the current timestamp. Hence, we choose to buffer all the required past features on-chip until the end of their lifetime, while (re)fetching the weight data from memory when needed. For Cross3D, the temporal-dimension kernel size is 5, which means $[5+4\times(dilation-1)]\times feature\_size$ data to be buffered on-chip for the output of each intermediate causal layer. 
    \item \textit{The frame rate}: The required memory bandwidth and arithmetic throughput are scaled to support real-time operation on the incoming data samples. More precisely, when denoting the computations necessary for one SRP feature map as one ``frame'', the system needs to handle $\frac{fs}{K\times (1-overlapping\_ratio)}$ frames per second.
\end{enumerate}

Although this is only a very naive implementation, it enables the ablation study on all variants of Cross3D proposed in this section. These experiment outcomes will be discussed in Section \ref{section_experiments}. 

\section{Experiments}\label{section_experiments}
In this section, we conduct ablation experiments to quantify the benefits of the proposed LC-SRP-Edge and Cross3D-Edge in Section \ref{section_methodologies}.
We first introduce the dataset specifications in Section \ref{section_experiments_dataset} and present general experiment configurations in Section \ref{section_experiments_configuration}. 
Then, we list design parameters used for the ablation study in Section \ref{section_experiments_ablation_params}.
Finally, results and discussions are expatiated in Section \ref{section_experiments_algorithm} and \ref{section_experiments_hardware}. 

\subsection{Datasets}\label{section_experiments_dataset}
Both synthetic (for training and testing) as well as real recorded (for testing) datasets are used in this work.

\subsubsection{Synthesized Dataset}\label{section_experiments_dataset_syn}\,

We use the dataset simulator from the Cross3D project \cite{diaz2020cross3dproject}) to train and test the Cross3D and Cross3D-Edge, for its ability to generate acoustic environments with widely varying characteristics. The Cross3D simulator is a highly-configurable runtime simulator for indoor acoustic scenes considering noise and reverberation levels. In this paper, we are targeting single-moving-source scenarios.

Similar to the original Cross3D project, we take the human voice dataset LibriSpeech \cite{panayotov2015librispeech} as the dry clean audio, including the ``/train-clean-100/'' folder as the training source and the ``/test-clean/'' folder as the testing source. As indicated in Section \ref{section_algorithm_system_overview} and Fig. \ref{fig:cross3d-workflow}, we assume the existence of a voice activity detection (VAD) module in our workflow. That is, the LibriSpeech data is preprocessed and labeled with timestamps for human-speech snippets, serving as ground-truth information for the evaluation stage.

As shown in Fig. \ref{fig:cross3d-workflow}, the source signal sequence is synthesized by a randomly selected ``Acoustic Scene'', including a random selected set for the following parameters:

\begin{figure*}[h]
    \centering
    \setlength{\abovecaptionskip}{0.cm}
    \includegraphics[width=\linewidth]{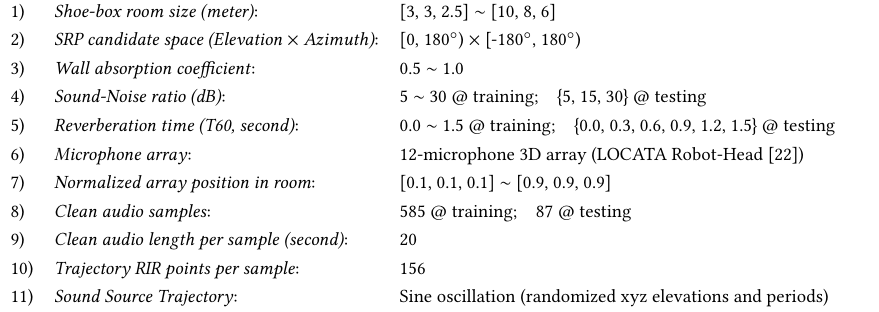}
\end{figure*}

Each time one dry clean audio snippet is fetched, a new ``Acoustic Scene'' is created with a re-selection of all parameters. Testing is performed on a partly fixed parameter set (e.g. specific SNR or T60) in line with the specifically targeted case studies.

Compared to the evaluation of the original literature, we make the following improvements:
\begin{enumerate}
    \item \textit{The range of microphone array position}: We change the range of this normalized position from [0.1, 0.1, 0.1] $\sim$ [0.9, 0.9, 0.5] to [0.1, 0.1, 0.1] $\sim$ [0.9, 0.9, 0.9] to cover most of the cases in the room space.
    \item \textit{The control of relative source distance:} We discard trajectories whose minimal distance to the microphone array is less than 1.0 meter, as a guarantee for the far-field propagation assumption (i.e. plain wave-front towards each microphone in the array) for SRP-PHAT computation.
    \item \textit{The static testing scenes}: We use static random number generators for the testing-stage Acoustic Scene construction, to ensure a fair comparison across all different models.
\end{enumerate}

We validate the benefits of these improvements with the SSL accuracy in Section \ref{section_experiments_algorithm} and \ref{section_sota_comparison}.
The pretrained models provided by the original codebase \cite{diaz2020cross3dproject} are considered as the \textbf{Cross3D(Baseline)} reference.
Note that for all experiments involving pretrained models, the aforementioned customized improvements are disabled, in order to avoid potential accuracy losses from transfer learning.

\subsubsection{Real Dataset}\label{section_experiments_dataset_real}\,

We also use the LOCATA dataset \cite{evers2020locata} to test and compare our models. The LOCATA data corpus is a real-world recorded dataset built for sound source localization and tracking. We pick \textit{Task1} and \textit{Task3} in the LOCATA development set for the test, which is built as one static source recorded by the static array and one moving source recorded by the static array, respectively. For the Robot-Head microphone array used in the Cross3D simulation, LOCATA provides 3 recordings for each task.

The dataset is recorded in a room of size [7.1, 9.8, 3.0] meter with T60$\approx$0.55 s, which is covered by the range of our synthesized Acoustic Scenes. Hence, it is reasonable to directly use the LOCATA tasks for testing, with our algorithms trained on the synthesized dataset. The result of this test comes with state-of-the-art comparisons in Section \ref{section_sota_comparison}.

\subsection{Experiment Configurations}\label{section_experiments_configuration}
\rebuttal{All software experiments are carried out on the NVIDIA RTX 2080Ti GPU platform, Python 3.8.8, and PyTorch 1.7.1, along with prerequisites from the original Cross3D repository \cite{diaz2020cross3dproject}. In terms of real hardware latency evaluation, we choose Raspberry Pi 4B as a representative embedded hardware platform, with the TVM toolchain \cite{chen2018tvm} for device-based algorithm optimization and deployment.}

The neural network model is trained with following hyper-parameters: maximal epoch size of 80, early-stopping patience of 22, initial learning rate of 0.0005, batch size of 5, and a fixed beginning SNR of 30dB. An explicit overwrite is performed at epoch-40, including updates of the learning rate to 0.0001, batch size to 10, and SNR to random range of 5$\sim$30 dB. During the training, a PyTorch Adam optimizer \cite{kingma2014adam,loshchilov2017decoupled} is applied to adjust the learning rate. The SSL accuracy is computed with the root-mean-square angular error (\textbf{RMSAE}) metric on the azimuth-elevation DOAs, while the loss function is based on their normalized equivalents in the Cartesian coordinate system. Besides, the training-stage RMSAE score is applied to the early-stopping module.

To enhance model robustness, both speech and non-speech snippets are taken into account during the training stage to calculate the loss function and SSL accuracy. 
In the evaluation stage of this ablation section with synthetic data, we focus on the SSL accuracy on our target, i.e. human speech snippets. That is, only ``no-silence'' RMSAE scores are selected with the help of VAD reference indices.
In the evaluation stage with real recorded data from the LOCATA dataset (Section \ref{section_sota_comparison}), we use mean angular error (\textbf{MAE}) on the entire recordings to be consistent with the SoTA research on this dataset.   

In addition, the experiments in this paper are computed on the 32-bit floating-point datatype. Accordingly, the hardware performance metric in Section \ref{section_experiments_hardware} is based on floating-point operations per second (FLOPS).

\begin{figure*}[t]
    \centering
    \setlength{\abovecaptionskip}{0.cm}
    \includegraphics[width=\linewidth]{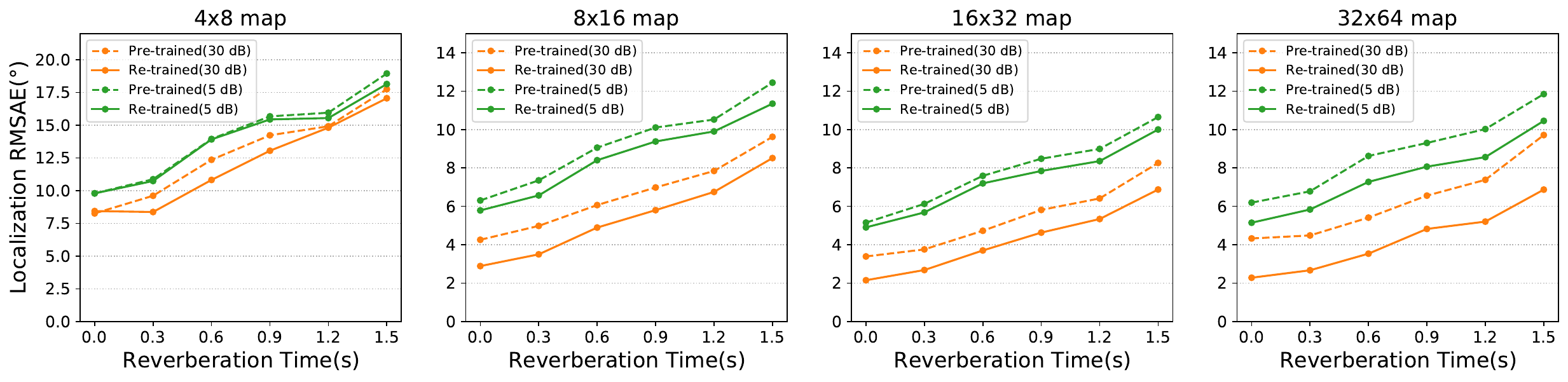}
    \caption{The localization RMSAE scores (the smaller, the better) of the pre-trained (Diaz-Guerra,2020) and our re-trained \textbf{Cross3D(Baseline)} model. The TD-SRP is used here as the input feature for both models. 
    }
    \label{fig:monotonic-rmsae-all-cases-ablation}
\end{figure*}

\subsection{Ablation Experiment Parameters}\label{section_experiments_ablation_params}
In line with Section \ref{section_methodologies}, we hereby list the algorithms involved in ablation experiments, along with the annotation for 4 representative corner cases and several customized design parameters, including the convolution-layer channel size and the source signal re-sampling.

We choose these ablation corner cases (\textbf{L}ow/\textbf{H}igh) to simplify the illustration and discussion of the SSL accuracy variation trend among the total 18 testing cases (3 SNRs and 6 T60s). 
Besides, with the original LibriSpeech's 16kHz audio, we leverage the Python library \textit{librosa} \cite{mcfee2015librosa} to enable the source signal re-sampling in point-(5). The parameter set of sampling rate is chosen to preserve basic characteristics of human speech, as the even lower rate severely damages the SSL accuracy in our investigation. Moreover, to keep the same temporal perception of the algorithm with the original \cite{diaz2020robust}, the parameter \bm{$K$} for Fourier transform is scaled proportionally to the sampling rate \bm{$fs$} in these customized cases, i.e. $[K=4096, fs=16000]$ versus $[K=2048, fs=8000]$.
The detailed parameters are as follows:

\vspace{1ex}
\begin{figure*}[h]
    \centering
    \setlength{\abovecaptionskip}{0.cm}
    \includegraphics[width=\linewidth]{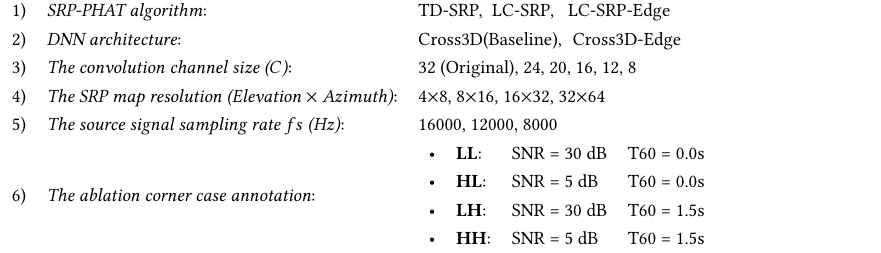}
\end{figure*}
\vspace{-2ex}

\subsection{Ablation Study on Algorithm Aspects}\label{section_experiments_algorithm}
This ablation experiment will study the algorithmic impact of different design parameters on Cross3D's methodologies from Section \ref{section_methodologies_ours}. 

\vspace{-2ex}
\subsubsection{Comparison of different SRP-PHAT methods}\label{section_experiments_algorithm_srp}\,

As a beginning experiment, we compare the pre-trained \cite{diaz2020cross3dproject} and our re-trained Cross3D(Baseline) models to validate our data augmentation methods (customized dataset and training configurations) in Section \ref{section_experiments_dataset} and \ref{section_experiments_configuration}. From the results in Fig. \ref{fig:monotonic-rmsae-all-cases-ablation}, we can conclude that the re-trained Cross3D(Baseline) outperforms the original baseline model across all acoustic environments and SRP-PHAT resolutions.
Higher-resolution cases, such as the 16$\times$32 and 32$\times$64, benefit more from this optimization, which is reasonable because more detailed SRP-PHAT candidate space is applied. On the contrary, the 4$\times$8 scenario barely shows the difference between the two cases, indicating this SRP resolution is too coarse-grained to depict the acoustic field. 

\begin{figure}[t]
    \centering
    \setlength{\abovecaptionskip}{0.cm}
    \includegraphics[width=.8\linewidth]{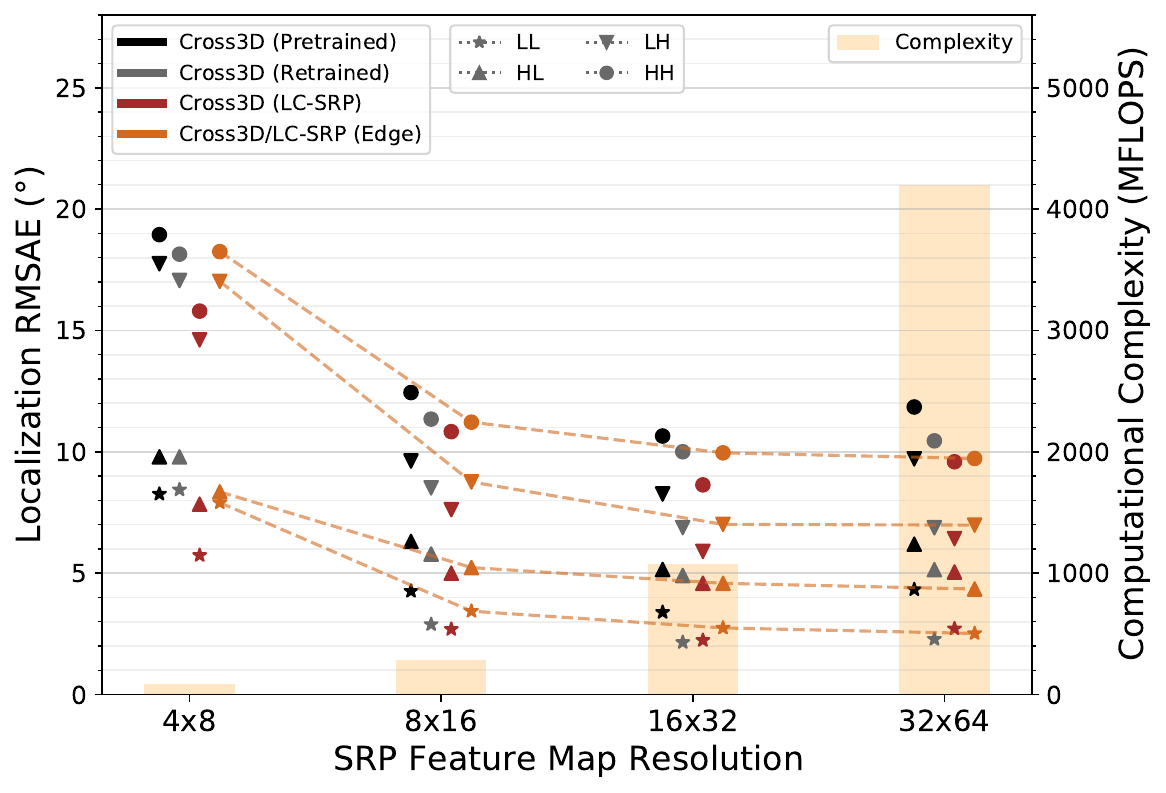}
    \caption{The comparison of Cross3D's localization RMSAE and computational complexity per second with different models: 1) The pre-trained Cross3D(Baseline); 2) The re-trained Cross3D(Baseline); 3) The Cross3D(Baseline) with LC-SRP feature map; 4) The proposed Cross3D-Edge with LC-SRP-Edge feature map. All cases work on the design parameter of $fs=16kHz, C=32$.}
    \label{fig:complexity-accuracy-resolution}
\end{figure}

One step further, we introduce the combination of different SRP-PHAT algorithms and Cross3D models towards our proposed model. Starting from the pre-trained and re-trained Cross3D(Baseline) model, we add two models which utilize the LC-SRP feature map and the Cross3D-Edge with LC-SRP-Edge structure, respectively. On top of the Cross3D(Baseline) structure, Cross3D-Edge introduces the usage of depth-wise layers. At this stage, all DNN models share the same design parameter $fs=16kHz, C=32$. Shown in Fig. \ref{fig:complexity-accuracy-resolution}, the localization accuracy is reported on 4 ablation corner cases as defined in Fig. \ref{section_experiments_ablation_params}.

Firstly, Cross3D(Baseline) with LC-SRP feature map replacement (brown dots) improves the localization accuracy one step further. For harsh environments, i.e. high T60s and low SNRs, this trend is clear across all SRP resolutions. This testifies the effectiveness of LC-SRP's interpolation method to reserve more signal information than the TD-SRP in Cross3D(Baseline). Besides, for easier scenes with low noise and reverberation levels, the aforementioned accuracy saturation phenomenon (Section \ref{section_methodologies_original_cross3D_bottleneck}) is again manifested, especially on the ``star'' cases which show mostly the same RMSAEs at 4$\times$8, 16$\times$32, and 32$\times$64 scenarios. 

Secondly, our target models are also evaluated, which leverages the Cross3D-Edge with LC-SRP-Edge. Basically, LC-SRP-Edge is mathematically equivalent to the original LC-SRP. Hence, the differences in SSL performance should only result from the modifications on DNN architecture. Shown with orange dots in Fig. \ref{fig:complexity-accuracy-resolution}, the RMSAEs compromise a bit compared to Cross3D(LC-SRP). It is reasonable because the DNN weights amount decreases by 50$\sim$75\% (Table \ref{tab:hardware-overhead-ablation}) after the usage of depth-wise layers. However, compared with the Cross3D(Retrained) scenario, our Cross3D-Edge retains competitiveness, with the same-level accuracy at the harsh \textbf{HH} and \textbf{LH} corners. Besides, minor accuracy diversity lies in \textbf{LL} and \textbf{LH} corners. We attribute such result to the turbulence in DNN training with the random training dataset (Section \ref{section_experiments_dataset}) unique to each training attempt. To conclude, we reckon our Cross3D-Edge structure to be a successful Cross3D variant. The benefits are discussed further in Section \ref{section_experiments_hardware} with the help of hardware metrics.

Thirdly, we also plot the computational complexity per inference in Fig. \ref{fig:complexity-accuracy-resolution}. Regarding to Fig. \ref{fig:workload-distribution-original}, all variants in this section report almost identical complexity for sharing the same dominant \textit{Cross\_Conv} blocks ($C=32$). In line with Table \ref{tab:original_cross3d_efficiency}, the accuracy saturation and complexity explosion at higher SRP-PHAT resolutions result in a tradeoff point at the 8$\times$16 SRP scenario. In the following ablation studies, we intend to focus on 8$\times$16 cases. Later in Section \ref{section_sota_comparison}, we will extrapolate the conclusions obtained from 8$\times$16 resolution to 16$\times$32 and 32$\times$64 to prove the generality.

\vspace{-2ex}
\subsubsection{Impact of different algorithmic parameters}\label{section_experiments_algorithm_param}\,

\begin{figure}[t]
    \centering
    \setlength{\abovecaptionskip}{0.cm}
    \includegraphics[width=.8\linewidth]{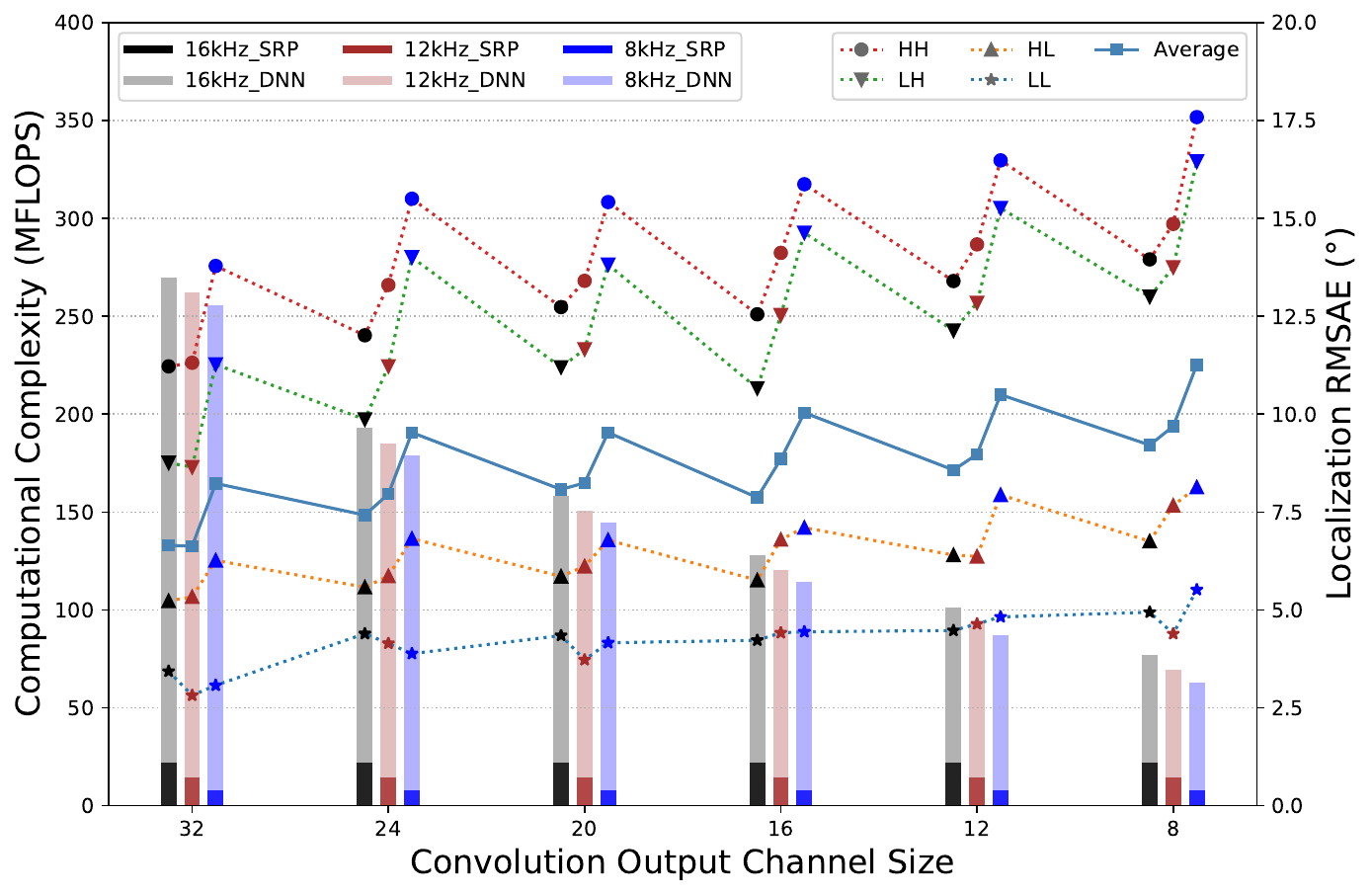}
    \caption{The ablation study of the proposed 8$\times$16 Cross3D-Edge's computational complexity per second and localization accuracy at different input audio qualities and convolution output channel sizes on average (18 noise-reverberation scenarios) and corner cases (\textbf{LL}, \textbf{HL}, \textbf{LH}, \textbf{HH}). The ablation parameter set is \bm{$fs$}: \{16000, 12000, 8000\} and \bm{$C$}: \{32, 24, 20, 16, 12, 8\}. }
    \label{fig:complexity-accuracy-channel-kfs}
\end{figure}

In this subsection, we evaluate the impact of our customized algorithmic parameters on localization errors, including the source signal sampling rate \bm{$fs$} and the convolution output channel size \bm{$C$}. The considered parameter sets are mentioned in Section \ref{section_experiments_ablation_params}. 
As stated in Section \ref{section_experiments_algorithm_srp}, we switch to study the proposed Cross3D-Edge with LC-SRP-Edge features on 8$\times$16 SRP-PHAT resolution.
Fig. \ref{fig:complexity-accuracy-channel-kfs} shows the corner-case-wise and average SSL accuracy in function of RMSAE, while varying \bm{$fs$} and \bm{$C$}. The corresponding computational complexity of LC-SRP-Edge and Cross3D-Edge for one inference is also plotted aside.
Note that mild accuracy turbulence also occurs in some scenarios (e.g. \textbf{LL} @ $C=32,\,24,\,20$) as in Fig. \ref{fig:complexity-accuracy-resolution}. We attribute this minor difference to the random and unique training dataset.

Among these results, one can first of all notice the monotonic decay of localization accuracy (i.e. increasing RMSAEs). This is expected, as for smaller \bm{$fs$} and \bm{$C$}, the SSL problem switches from ideal acoustic environments to harsher ones (i.e. lower-quality source signals) and the model's trainable parameters decrease when the CNN channel size becomes even smaller. 
In return, the DNN and SRP complexity decreases almost proportionally to these two design parameters. 

Generally, the two design parameters \bm{$C$} and \bm{$fs$} impact the localization accuracy differently when shrinking the volume of algorithm. 
On the one hand, the conv-layer channel size (\bm{$C$}) controls the volume of the dominant neural network blocks, implicitly affecting the DNN generality. Considering the most distant \bm{$C=32$} and \bm{$C=8$} groups, the average RMSAE score only decreases by 38.5\%, 46.3\%, and 36.8\%, while the complexity is reduced by 69.4\%, 72.4\%, and 75.8\%, respectively. Meanwhile, the localization error is still within the SRP grid threshold (22.5$^\circ$ @ 8$\times$16), which is much better than traditional methods discussed in the original literature \cite{diaz2020robust}. 

On the other hand, the source signal quality (\bm{$fs$}) impacts SSL RMSAEs greatly. One can see in Fig. \ref{fig:complexity-accuracy-channel-kfs} that the largest model with worst-quality source (\bm{$C=32, fs=8kHz$}) produces similar accuracy to the smallest model with original-quality source (\bm{$C=8, fs=16kHz$}), while the former consumes $>300\%$ more computations compared to the latter, showing the extra DNN efforts to handle a low-quality source. On the same DNN version, the localization accuracy deteriorates 1x$\sim$4x faster in \textit{12kHz->8kHz} cases than the \textit{16kHz->12kHz}. This is reasonable as the \bm{$fs=8kHz$} cases are already the critical sampling rate for human voice without sibilance, indicating the great loss of high-frequency information. Meanwhile, this outcome also suggests the diminishing marginal efficiency to pursue higher \bm{$fs$} (e.g. 44.1kHz, 48kHz, or higher) in human-voice sound source localization. Considering the impact of \bm{$K-fs$} parameter on complexity in Table \ref{tab:hardware-overhead-summary}, it would be an interesting future direction to study the necessary minimal sampling rate for localizing specific types of sound other than speech towards the lower computational cost. 
However, restricted by the LibriSpeech's 16kHz recording, which is much lower than other datasets such as the 48kHz TAU Spatial Sound Events 2019 dataset \cite{adavanne2019multi}, we are not able to conduct those experiments in this paper. 

\begin{table}[t]
    \centering
    \caption{Result comparison of the customized tradeoff metric (\bm{$Complexity \times RMSAE$}, the smaller the better), computed with the overall complexity and the average RMSAE from Fig. \ref{fig:complexity-accuracy-channel-kfs}. The minimal value is marked in bold text for each parameter $C$. As a reference, the score for the pretrained Cross3D(Baseline) \cite{diaz2020robust} is \textit{\textbf{2.09}}.
    }
    \label{tab:customized-efficiency-tradeoff-metric-cfs}
    \includegraphics[width=\linewidth]{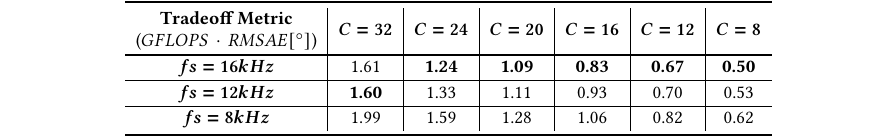}
\end{table}

To sum up, our scaling scheme of design parameter \bm{$C$} on Cross3D-Edge benefits the algorithm complexity while retaining good robustness against noise and reverberation cases. 
The assessment across different \bm{$fs$} manifests the importance of source signal quality. 
To compile different cases into one, we introduce a new efficiency metric as $Complexity \times RMSAE$ as a reference. Shown in Table \ref{tab:customized-efficiency-tradeoff-metric-cfs}, it is a tradeoff indicator to provide an intuitive view of the overall influence of our design parameter study, in which lower values are aimed for. 
Under this metric, the best tradeoff points are achieved at $fs=16kHz$, except for a minor fallback for $C=32$. This agrees with our previous discussion on the SSL model's sensitivity to source signal quality. 
Compared with Cross3D(Baseline) \cite{diaz2020robust} which scores \textbf{2.09}, the benefit of Cross3D-Edge and LC-SRP-Edge as a better accuracy-complexity balance is clearly manifested.

Although our metric equally values the accuracy and complexity here, developers can also choose between these design parameters by their specific design focus.

For the next sections, we select 3 representative size among the above versions. Named as \textit{Cross3D-Edge-Large (\textbf{EL})}, \textit{Cross3D-Edge-Medium (\textbf{EM})}, and \textit{Cross3D-Edge-Small (\textbf{ES})}, the neural networks inside are with parameter \bm{$C=32$}, \bm{$C=16$}, and \bm{$C=8$}, respectively. As the impact of \bm{$fs$} is minor on the complexity in Fig. \ref{fig:complexity-accuracy-channel-kfs}, all 3 versions use \bm{$fs=16kHz$} SRP for better localization accuracy. 
For instance, the selection among these versions could be accuracy-oriented (\textit{\textbf{EL}}), footprint-sensitive (\textit{\textbf{ES}}), or the tradeoff (\textit{\textbf{EM}}).

\begin{figure}[t]
    \centering
    \includegraphics[width=0.8\linewidth]{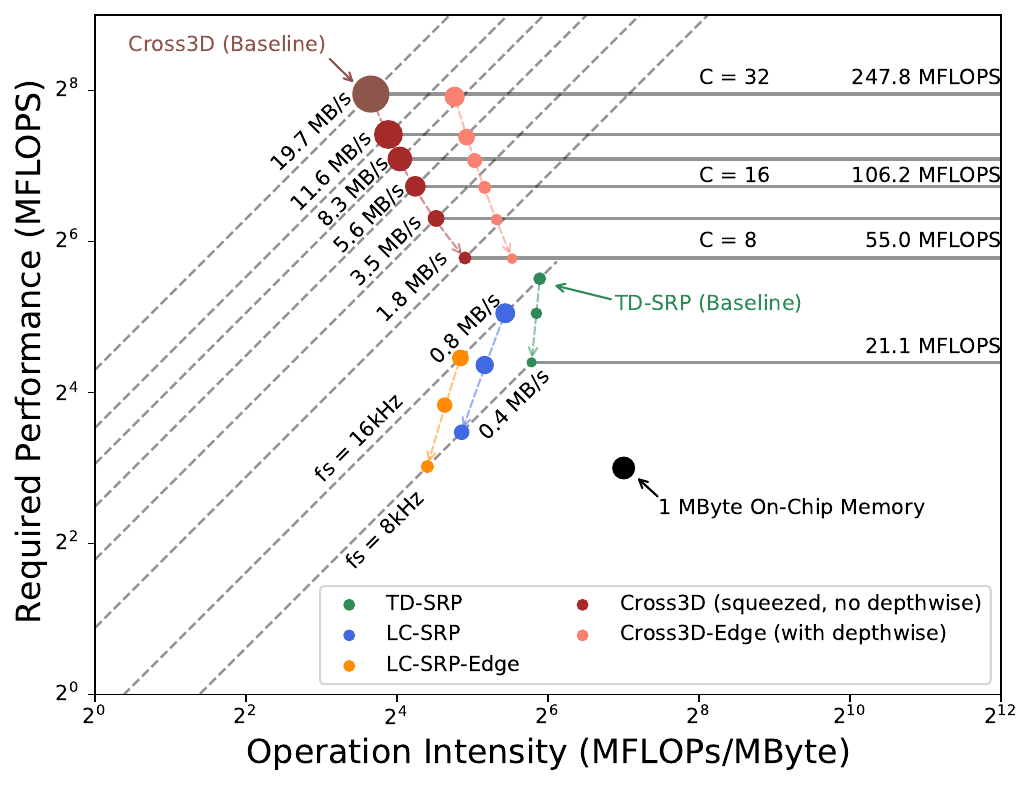}
    \caption{Roofline analysis of the hardware overhead per second to implement the algorithm versions in Section \ref{section_experiments_algorithm}. The DNN structures are based on our Cross3D-Edge (Fig. \ref{fig:cross3d-methodologies} (b)), with and without the depth-wise layers. The diameter of data points illustrates the required on-chip memory of the related version. The arrows in figure display the design parameter differences from high-end to low-end, including \bm{$C\in\{32, 24, 20, 16, 12, 8\}$} for Cross3D series and \bm{$fs\in\{16kHz, 12kHz, 8kHz\}$} for SRP series. The prefix of our metrics is decimal, i.e. $M=1e6$ for MFLOPs, MByte, and MB/s. 
    }
    \label{fig:roofline_fmap}
\end{figure}

\subsection{Ablation Study on Hardware Aspects}\label{section_experiments_hardware}
Along with the algorithmic ablations, this subsection reflects on the hardware-level benefits of the Cross3D-Edge and LC-SRP-Edge approach for execution in extreme edge devices. Based on the analysis and metrics in Section \ref{section_methodologies_overhead}, we leverage the roofline model \cite{williams2009roofline} to show the differences between model versions in terms of computational complexity, operation intensity, on-chip memory overhead, and the required memory bandwidth. The resulting per-second overhead is summarized in Fig. \ref{fig:roofline_fmap}. In our setup, the neural network dominates hardware overhead. However, this could be different with other source signal and microphone array properties impacting the SRP overhead summarized in Table \ref{tab:hardware-overhead-summary}.

The proposed Cross3D-Edge network benefits from the squeezed ``\bm{$C$}'' parameters and the depth-wise layers to save computations and memory overhead, respectively. As discussed in Section \ref{section_methodologies_original_cross3D_bottleneck}, Cross3D(Baseline) model leads to large hardware overhead, including 247.8 MFLOPS performance, 19.7 MByte/s bandwidth, and 2.83 MByte on-chip memory. 
The effects of reducing parameter \bm{$C$} is shown by the intermediate Cross3D-Edge (no depth-wise) scenario, where the computational complexity is reduced to a great extent. At the minimal \bm{$C=8$}, these overheads can be controlled to 55.0 MFLOPS performance, 1.8 MByte/s bandwidth, and 0.273 MByte on-chip memory.
Meanwhile, in Cross3D-Edge (with depth-wise), the usage of depth-wise layers further reduces memory overhead, moving the roofline points to the right. For example, our \textit{Cross3D-Edge-Large} model (\bm{$C=32$}) only requires 8.9 MByte/s bandwidth and 0.76 MByte on-chip memory, which is only 45.1\% and 26.8\% of the Cross3D(Baseline).

Focusing on the SRP calculations, the switch from TD-SRP to LC-SRP/LC-SRP-Edge mainly contributes to the reduction of computational complexity. On \bm{$fs=16kHz$} source, these SRP-PHAT computation consumes 45.49 MFLOPS, 33.13 MFLOPS, and 21.97 MFLOPS per second, respectively. Mentioned in Section \ref{section_methodologies}, LC-SRP saves computation by only computing Fourier transform on necessary sample points. However, LC-SRP needs more memory (larger dot size in Fig. \ref{fig:roofline_fmap}) to store the sinc-interpolation coefficients. For signals of \bm{$fs=16kHz$}, \bm{$fs=12kHz$}, and \bm{$fs=8kHz$}, TD-SRP requires 0.263 MByte, 0.205 MByte, and 0.148 MByte of on-chip memory space, while LC-SRP needs 0.737 MByte, 0.611 MByte, and 0.419 MByte, accordingly. After our optimization in \textit{LC-SRP-Edge}, such overhead is reduced to 0.534 MByte (72.5\%), 0.443 MByte (72.5\%), and 0.318 MByte (75.8\%), respectively. Meanwhile, the proposed \textit{LC-SRP-Edge} saves computation by 33.7\%, 30.8\%, and 27.0\% with reference to Eq. \ref{eq:complexity-lcsrp-edge}.

In Fig. \ref{fig:complexity-accuracy-channel-kfs} and \ref{fig:roofline_fmap}, we only discuss the 8$\times$16 resolution. However, the proposed optimizations on SRP and DNN structures can be exploited across all resolutions. Hence, we extrapolate the evaluation to multiple resolution cases in Table \ref{tab:hardware-overhead-ablation}. The computational complexity drops by 10.32\%, 56.72\%, and 73.71\% on average with the proposed \textit{Cross3D-Edge} and \textit{LC-SRP-Edge}. In addition, the parameter amount shows even greater advantages towards resource-constrained edge devices. The the average weight volume reduction is 59.77\%, 86.74\%, and 94.66\% on the three \textit{Cross3D-Edge} series.

\begin{table}[t]
    \centering
    \setlength{\abovecaptionskip}{0.cm}
    \setlength{\belowcaptionskip}{0.cm}
    \renewcommand\arraystretch{1.08}
    \caption{The hardware overhead comparison between the Cross3D(Baseline) \cite{diaz2020robust} and the proposed \textit{Cross3D-Edge} series (\bm{$C=32, 16, 8$}) across 4 resolution scenarios. The computational complexity \rebuttal{(\textbf{Op \#}, in MFLOPs)} is calculated for each SRP frame, including the SRP and DNN computation. The parameter amount \rebuttal{(\textbf{Param \#})} only considers the DNN weights as SRP coefficients are cached in on-chip memory (Section \ref{section_methodologies_overhead}).}
    \label{tab:hardware-overhead-ablation}
    \includegraphics[width=\linewidth]{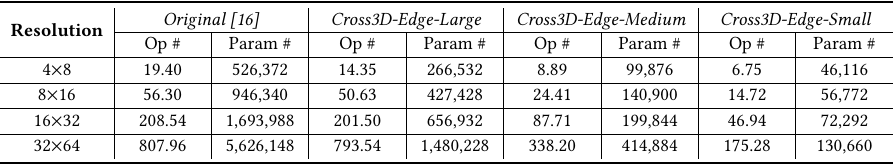}
\end{table}

\begin{table}[t]
    \centering
    \setlength{\abovecaptionskip}{0.cm}
    \setlength{\belowcaptionskip}{0.cm}
    \renewcommand\arraystretch{1.08}
    \caption{\rebuttal{Real hardware processing latency comparisons (in \textit{\textbf{milliseconds}}) of SRP computation and DNN inference, between the proposed approach and baselines \cite{diaz2020cross3dproject,diaz2020robust, dietzen2020low}. The workload is set to 1 windowed frame of microphone-array signals. The implementation and evaluation is carried out on 1 Raspberry Pi 4B board with the help of the TVM toolchain \cite{chen2018tvm}.}}
    \label{tab:hardware-deployment-evaluation}
    \includegraphics[width=\linewidth]{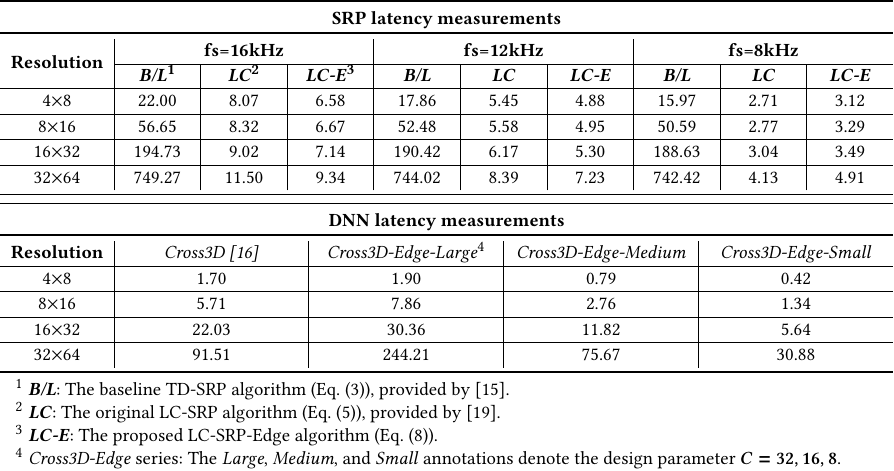}
\end{table}

\rebuttal{
To verify the efficacy of our proposed optimizations, we evaluate the runtime execution latency on a representative embedded device (Raspberry Pi 4B, with 4-core 64bit Cortex-A72 @ 1.5GHz processor, LPDDR4-3200 memory access and 1 MiB L2 Cache). We leverage the TVM toolchain \cite{chen2018tvm} to obtain optimal implementations on the target device. Table \ref{tab:hardware-deployment-evaluation} summarizes the latency comparison over all scenarios, where several positive conclusions can be drawn:
Firstly, based on the analysis in Section \ref{section_experiments_algorithm} and \ref{section_experiments_hardware}, a typical accuracy-overhead tradeoff point is  \textit{Cross3D-Edge-Medium (\textbf{EM})} with 16kHz-sampled inputs and 8$\times$16 LC-SRP-Edge feature maps. The computation latency for such workload is 8.59 ms/frame (5.83 ms for SRP computation and 2.76 ms for DNN inference), enough to satisfy real-time processing requirements, enabling up to 116 frames per second.
Secondly, the LC-SRP algorithms show stable low-latency performance over all the sampling frequencies. On top of that, the proposed LC-SRP-Edge optimizations further reduce such latency at high quality recordings ($fs = 16kHz, 12kHz$) by 12\%$\sim$22\%. It is also feasible to rapidly calculate SRP when the resolution scales up. Compared to this, although the baseline TD-SRP algorithm can better benefit from hardware FFT kernels, the irregular memory access in Eq. \ref{eq:srp_phat_td}'s indexing procedure hinders the parallelism and leads to a rapid latency increase for higher SRP resolutions.
Thirdly, over most resolutions, the DNN inference latencies of Cross3d-Edge series are well-contained. The Cross3D-Edge-Large versions are slightly slower than the original baseline due to the complexity of depth-wise convolutions. For \textbf{\textit{EM}} and \textbf{\textit{ES}} models, the inference latency is saved by 42\% and 73\%, which is in line with Fig. \ref{fig:roofline_fmap}. However, the proposed Edge models also hold the advantage of reduced DNN parameters (see Table \ref{tab:hardware-overhead-ablation}), which is not revealed on the resource-abundant Raspberry Pi 4 platform. Meanwhile, the 32$\times$64 DNN tends to exceed the capability of this embedded platform from its overall latency explosion, while it is also proved to be an overkill of the mission in Fig. \ref{fig:complexity-accuracy-resolution}.
}

In short, this section elucidates the improvement of the hardware friendliness of the proposed \textit{Cross3D-Edge} and \textit{LC-SRP-Edge} algorithms. The optimization methodologies greatly reduce the hardware overhead in computational complexity, memory bandwidth, and on-chip memory size. This enables the deployment of the Cross3D model at extreme edge devices. 

\section{State-of-the-Art Comparison and Discussion}\label{section_sota_comparison}
To further validate the efficacy of the proposed optimizations and benchmark against the SotA, we test our \textit{Cross3D-Edge-Medium with \textit{LC-SRP-Edge}} architecture (\textit{\textbf{EM}}) on the extensively benchmarked real-world recorded data from the LOCATA corpus \cite{evers2020locata}. We focus on LOCATA \textit{Task1}, recorded with a static microphone array for a single static sound source, and \textit{Task3}, dealing with a single moving sound source. The results are summarized in Table \ref{tab:sota-comparison}, and compared to the original Cross3D model and several state-of-the-art works. 
    
\begin{table}[t]
    \centering
    \setlength{\abovecaptionskip}{0.cm}
    \setlength{\belowcaptionskip}{0.cm}
    \caption{The localization accuracy and comparison on \textit{Task1} and \textit{Task3} of the LOCATA dataset. The comparison involves the Cross3D baseline \cite{diaz2020robust}, SELDnet \cite{adavanne2018sound}, Grumiaux\,(2021)\cite{grumiaux2021improved}, and Perotin\,(2018) \cite{perotin2018crnn}. The metric of localization is mean angular error (MAE) in degrees. The neural network parameter amount is also reported. The relative best cases are marked in bold.}
    \label{tab:sota-comparison}
    \includegraphics[width=\linewidth]{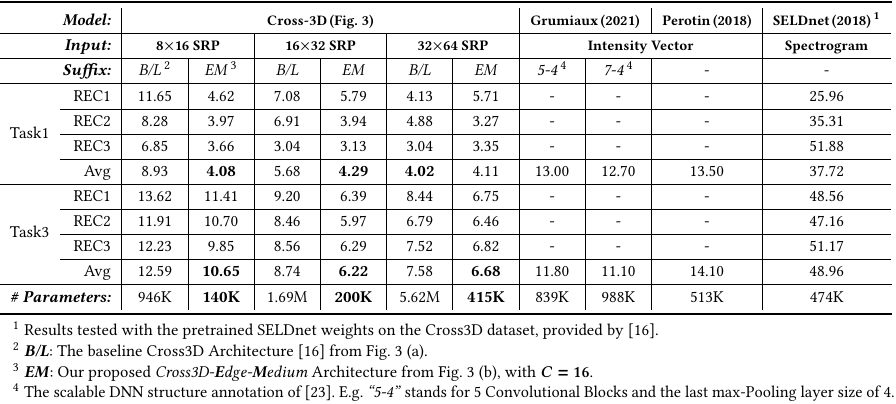}
\end{table}

The results show that the \textit{EM} model surpasses the Cross3D baseline in almost every scenario by 11.8\%$\sim$54.3\%. The only fallback is at the 32$\times$64 SRP for \textit{Task1}, with a negligible MAE difference of 2.2\%. Thanks to the proposed algorithmic optimizations (Section \ref{section_methodologies_ours}) and the dataset refinement (Section \ref{section_experiments_dataset_syn}), the proposed model attains better robustness and generality on unseen real-world recordings.

Moreover, the proposed model also outperforms the state-of-the-art research included in Table \ref{tab:sota-comparison}. On the one hand, the two intensity-vector-based models \cite{perotin2018crnn,grumiaux2021improved} come with larger model sizes, while not bringing better performance. They are originally designed to handle multiple moving sound sources with CRNN, which is an unsupported feature in Cross3D, but as reported in \cite{grumiaux2021improved}, their performance is almost the same between single-source and multi-source tasks on the LOCATA challenge. 
On the other hand, the SELDnet \cite{adavanne2018sound} is targeting single source localization tasks, yet does not succeed to generalize all the Acoustic Scenes during the training. The reported numbers are obtained by retraining \cite{diaz2020robust} on Cross3D's synthesized data, showing insufficient generalization capabilities and noise robustness.

Last but not least, assessing the parameter amount of all benchmarked algorithms further highlights the benefits of the proposed \textit{Edge} model series as an efficient solution for extreme edge deployment. The selected \textit{\textbf{EM}} model, along with other variants in Table \ref{tab:hardware-overhead-ablation}, enables much smaller neural networks than the state-of-the-art with comparable SSL accuracy. Moreover, note that \textit{Cross3D-Edge} is a pure-CNN solution, which makes the proposed method also more hardware-friendly in terms of parallelization for real-time execution, compared with other research with CRNN or even more complex structures.

\section{Conclusion and Future Work}\label{section_conclusion}
In this paper, we conduct optimizations on the computation of SRP-PHAT and Cross3D neural network towards low hardware footprints for extreme edge implementation. Based on the bottleneck analysis, hardware-friendly \textit{LC-SRP-Edge} and \textit{Cross3D-Edge} models are proposed. Ablation studies are carried out to further optimize and prove the efficacy of each modification. With the refinement in dataset generation and training configuration, the proposed algorithm outperforms the baseline method and state-of-the-art research both in terms of localization accuracy and hardware overhead. 
\rebuttal{We verify the end-to-end real-time processing capability of our proposed algorithms by the deployment and latency evaluation on an embedded device. The optimized model (\textit{Cross3D-Edge-Medium + LC-SRP-Edge}) requires only 127.1 MFLOPS computation, 3.71 MByte/s bandwidth, and 0.821 MByte on-chip memory in total, which results in 8.59 ms/frame overall latency on Raspberry Pi 4B. }

Based on the analysis in this paper, several interesting future directions can be explored: 1) The extrapolation of Cross3D's structure to support multiple sound sources, overlapping utterance, etc. 2) The adoption of SRP-PHAT-based localization models into the prevalent multi-head neural network structures on sound event localization, detection, and tracking (SELDT) missions. 3) The effective hardware-software co-design system for hardware-friendly SELDT solutions.

\begin{acks}
This project has received funding from the European Union’s Horizon 2020 research and innovation programme under the Marie Skłodowska-Curie grant agreement No. [956962].
\end{acks}

\bibliographystyle{ACM-Reference-Format}
\bibliography{bibli}

\end{document}